# Simple and robust element-free Galerkin method with interpolating shape functions for finite deformation elasticity

**George Bourantas[1], Benjamin F. Zwick[1], Grand Joldes[1], Adam Wittek[1], Karol Miller[1,2]**

*[1]Intelligent Systems for Medicine Laboratory, The University of Western Australia, Perth 6009, Western Australia, Australia*

*[2]Harvard Medical School, Boston, Massachusetts, USA*

**Abstract:** In this paper, we present a meshless method belonging to the family of element-free Galerkin (EFG) methods. The distinguishing feature of the presented meshless method is that it allows accurate enforcement of essential boundary conditions. The method uses total Lagrangian formulation with explicit time integration to facilitate code simplicity and robust computations in applications that involve large deformations and non-linear materials. We use a regularized weight function, which closely approximates the Kronecker delta, to generate interpolating shape functions. The imposition of the prescribed displacements on the boundary becomes as straightforward as in the finite element (FE) method. The effectiveness and accuracy of the proposed method is demonstrated using 3D numerical examples that include cylinder indentation by 70% of its initial height, and indentation of the brain.

## 1. Introduction

The finite element method (FEM) dominates the field of numerical analysis in non-linear solid mechanics. Nevertheless, it is well known that FEM has important deficiencies, both practical and theoretical. They include, among others, practical difficulty of quality finite element mesh generation and deterioration of accuracy or outright failure of computation when elements are excessively distorted due to large deformations. Therefore, other alternatives are called for.

Meshless methods have a very attractive advantage: there is no need for mesh generation. Instead, the spatial domain is represented by a set of points or nodes. For nonlinear problems, methods like meshless local Petrov-Galerkin (MLPG) [1, 2], smooth particle hydrodynamics [3], element-free Galerkin (EFG) [4] and strong form collocation methods [5, 6] have demonstrated their accuracy in several benchmark problems, in many cases superior to FE methods. Recent refinements of element-free Galerkin methods appear to be most promising, with comprehensive 3D non-linear problems solved effectively and accurately [7-9].

Unfortunately, current implementations of meshless methods are complicated and lack robustness, mostly because of difficulties arising from recalculation of shape functions (in updated Lagrangian implicit schemes), inadmissibility of certain nodal configurations (due to singularities of the moment matrix), application of essential boundary conditions (due to commonly used meshless shape functions being non-interpolating), and volumetric integration of non-polynomial integrands.

In this paper, we present a simple and robust EFG method. Total Lagrangian formulation with explicit time integration eliminates the need of multiple recalculation of shape functions and their derivatives. Modified moving least squares (MMLS) [10] render (for practical purposes) almost all nodal distributions admissible. The proposed improvement to MMLS renders the shape functions almost interpolating, allowing direct imposition of boundary conditions (as in FEM). Extremely simple volumetric integration using an automatically generated tetrahedral background integration grid (with no stringent quality requirements), while not resolving theoretical difficulties of integrating rational functions, provides sufficient accuracy in practice.

The paper is organized as follows: in Section 2 we briefly describe the components of our methodology adapted from our previous work [6, 9, 11], and explain improvements to the MMLS functions [10] that render them interpolating. In Section 3 we demonstrate convergence of our algorithm as well as its accuracy by comparing our results to analytical and finite element solutions. Section 4 contains two examples for which it would be very difficult to obtain solutions with any other numerical method currently available. Section 6 contains discussion and conclusions.

## 2. Numerical method

### *2.1 Total Lagrangian Explicit Dynamics (TLED)*

In the total Lagrangian (TL) formulation [12, 13] all calculations refer to the initial configuration of the analyzed continuum. Spatial derivatives are computed during the pre-processing stage, which eliminates the necessity of re-evaluating shape functions and their derivatives at every time step (as in the updated Lagrangian (UL) formulation).

After introducing the approximation of the displacement field using the modified moving least squares (MMLS) method (details in Section 2.2) into the weak form of governing equations of solid mechanics using the TL formulation [14], the global system of discretized equations describing the behavior of the analyzed continuum becomes

$$\mathbf{M}\,{}^{t}\ddot{\boldsymbol{u}} + {}^{t}\mathbf{F}_{int} = {}^{t}\mathbf{F}_{ext}\,, \tag{1}$$

where $\mathbf{M}$ is the constant mass matrix, ${}^{t}\ddot{\boldsymbol{u}}$ is the vector of nodal accelerations, ${}^{t}\mathbf{F}_{int}$ is the global nodal internal force vector and ${}^{t}\mathbf{F}_{ext}$ is the vector of externally applied forces at time $t$. The vector of internal nodal forces ${}^{t}\mathbf{F}_{int}$ is computed as:

$$ {}^{t}\mathbf{F}_{int} = \int_{V_0} {}_{0}^{t}\mathbf{X}\; {}_{0}^{t}\mathbf{B}_{L0}^{\mathrm{T}}\; {}_{0}^{t}\mathbf{S}\; \mathrm{d}V_0\,, \tag{2}$$

where ${}_{0}^{t}\mathbf{X}$ is the deformation gradient at time $t$, ${}_{0}^{t}\mathbf{S}$ is the second Piola-Kirchoff stress at time $t$, ${}_{0}^{t}\mathbf{B}_{L0}$ is the matrix of shape function derivatives and $V_0$ is the initial volume of the problem domain.

We apply explicit integration in time using the central difference method. Explicit time integration is a direct method, where nodal accelerations are integrated (using 2nd order accurate finite difference scheme) to obtain the displacements, and there is no need to assemble a global stiffness matrix. The time stepping scheme for the equation of motion can be expressed as:

$$ {}^{t+1}\boldsymbol{u} = \Delta t^2 \mathbf{M}^{-1}({}^{t}\mathbf{F}_{ext} - {}^{t}\mathbf{F}_{int}) + 2\,{}^{t}\boldsymbol{u} - {}^{t-1}\boldsymbol{u}, \tag{3}$$

where ${}^{t}\boldsymbol{u}$ is the displacement calculated at time $t$, $\mathbf{M}$ is the constant diagonal mass matrix and $\Delta t$ is the time step.

To accelerate the solution convergence towards its steady state, we apply the dynamic relaxation method presented in [15, 16]. The global system of discretized damped equations of motion is

$$\mathbf{M}\,{}^{t}\ddot{\boldsymbol{u}} + {}^{t}\mathbf{F}_{damp} + {}^{t}\mathbf{F}_{int} = {}^{t}\mathbf{F}_{ext}, \tag{4}$$

with $\mathbf{F}_{damp} = c\mathbf{M}^t\dot{\boldsymbol{u}}$ being the mass proportional damping force, and $c$ being the damping coefficient. The updated displacement field ${}^{t+1}\boldsymbol{u}$ is computed as

$$ {}^{t+1}\boldsymbol{u} = {}^{t}\boldsymbol{u} + \beta({}^{t}\boldsymbol{u} - {}^{t-1}\boldsymbol{u}) + \alpha\mathbf{M}^{-1}({}^{t}\mathbf{F}_{ext} - {}^{t}\mathbf{F}_{int}), \tag{5} $$

where $\alpha = 2\Delta t^2/(2 + c\Delta t)$ and $\beta = (2 - c\Delta t)/(2 + c\Delta t)$.

*2.2 Moving Least Squares (MLS) and Modified Moving Least Squares (MMLS) methods*

*Moving Least Squares (MLS)*

The technique summarized below is well established, see e.g. [1, 17]. To approximate the unknown field function $u(\boldsymbol{x})$ over a set of randomly distributed nodes $\{\boldsymbol{x}_i\}$, $i$=1,2,…,$n$, the moving least squares approximation $u^h(\boldsymbol{x})$ is defined as

$$ u^h(\boldsymbol{x}) = \sum_{j=1}^{m} p_j(\boldsymbol{x})a_j(\boldsymbol{x}) = \mathbf{p}^T(\boldsymbol{x})\boldsymbol{a}(\boldsymbol{x}), \tag{6} $$

where $\mathbf{p}^T(\boldsymbol{x}) = [p_1(\boldsymbol{x}), p_2(\boldsymbol{x}), \dots, p_m(\boldsymbol{x})\ ]$ are basis functions, $\boldsymbol{a}(\boldsymbol{x})$ is a vector containing the coefficients $a_j(\boldsymbol{x})$ $(j = 1,2, \dots, m)$, and $m$ is the number of terms in the basis $\mathbf{p}^T(\boldsymbol{x})$. The basis functions are usually chosen as the monomial bases. In the present study we chose the quadratic basis, which is defined as:

$$ \mathbf{p}^T(\boldsymbol{x}) = [1, x, y, z, x^2, y^2, z^2, xy, xz, yz]. \tag{7} $$

The coefficient vector $\boldsymbol{a}(\boldsymbol{x})$ is computed at any point $\boldsymbol{x}$ of the spatial domain by minimizing an error functional $J(\mathbf{x})$ defined based on the weighted least squares errors as

$$ \boldsymbol{J}(\boldsymbol{x}) = \sum_{i=1}^{n} w(\|\boldsymbol{x} - \boldsymbol{x}_i\|)(\mathbf{p}^T(\boldsymbol{x}_i)\boldsymbol{a}(\boldsymbol{x}) - u_i)^2 = (\mathbf{P}\boldsymbol{a}(\boldsymbol{x}) - \boldsymbol{u})^{\mathrm{T}}\mathbf{W}(\mathbf{P}\boldsymbol{a}(\boldsymbol{x}) - \boldsymbol{u}), \tag{8} $$

where $w(\|\boldsymbol{x} - \boldsymbol{x}_i\|)$ is a weight function associated with the node $i$ ($w(\|\boldsymbol{x} - \boldsymbol{x}_i\|) \geq 0$ for all $\boldsymbol{x}$ in the support domain of node), $\boldsymbol{x}_i$ the value of $\boldsymbol{x}$ at node $i$, $\boldsymbol{u} = u_i = (u_1, u_2, \dots, u_n)$ are the

values (often called fictitious values of the approximation) of the unknown field $u(\boldsymbol{x})$ at nodes $\boldsymbol{x}_i$, and $n$ the number of nodes in the support domain of $\boldsymbol{x}$. The matrices **P** and **W** in Eq. (8) are computed as

$$\mathbf{P} = \begin{bmatrix} p_1(\boldsymbol{x}_1) & p_1(\boldsymbol{x}_2) & \cdots & p_1(\boldsymbol{x}_n) \\ p_2(\boldsymbol{x}_1) & p_2(\boldsymbol{x}_2) & \cdots & p_2(\boldsymbol{x}_n) \\ \vdots & \vdots & \ddots & \vdots \\ p_m(\boldsymbol{x}_1) & p_m(\boldsymbol{x}_2) & \cdots & p_m(\boldsymbol{x}_n) \end{bmatrix}_{m\times n} \tag{9}$$

$$\mathbf{W} = \begin{bmatrix} w_1(\boldsymbol{x}) & 0 & \cdots & 0 \\ 0 & w_2(\boldsymbol{x}) & \cdots & 0 \\ \vdots & \vdots & \ddots & \vdots \\ 0 & \cdots & 0 & w_n(\boldsymbol{x}) \end{bmatrix}_{n\times n} . \tag{10}$$

To compute the coefficient vector $\boldsymbol{a}(\boldsymbol{x})$, we minimize the functional $J(\boldsymbol{x})$ with respect to the vector components $a_j(\boldsymbol{x})$ [18]

$$\frac{\partial \boldsymbol{J}}{\partial \boldsymbol{a}} = \boldsymbol{A}(\boldsymbol{x})\boldsymbol{a}(\boldsymbol{x}) - \boldsymbol{B}(\boldsymbol{x})\boldsymbol{u} = 0, \tag{11}$$

where

$$\boldsymbol{A}(\boldsymbol{x}) = \mathbf{PWP}^{\mathrm{T}} = \sum_{i=1}^{n} w_i(\boldsymbol{x})\mathbf{p}(\boldsymbol{x}_i)\mathbf{p}^{\mathrm{T}}(\boldsymbol{x}_i) \tag{12}$$

$$\boldsymbol{B}(\boldsymbol{x}) = \mathbf{PW} = \big(w_1(\boldsymbol{x})\mathbf{p}(\boldsymbol{x}_1), w_2(\boldsymbol{x})\mathbf{p}(\boldsymbol{x}_2), \dots, w_n(\boldsymbol{x})\mathbf{p}(\boldsymbol{x}_n)\big) \tag{13}$$

If the moment matrix $\boldsymbol{A}(\boldsymbol{x})$ is non-singular, from Eq. (11) we can obtain

$$\boldsymbol{a}(\boldsymbol{x}) = \boldsymbol{A}^{-1}(\boldsymbol{x})\boldsymbol{B}(\boldsymbol{x})\boldsymbol{u} . \tag{14}$$

By substituting $\boldsymbol{a}(\boldsymbol{x})$ into Eq. (6), we obtain for the local approximation $u^h(\boldsymbol{x})$

$$u^h(\boldsymbol{x}) = \Phi(\boldsymbol{x})\boldsymbol{u} = \sum_{i=1}^{n} \phi_i(\boldsymbol{x})u_i \tag{15}$$

with $\Phi(\boldsymbol{x}) = \big(\phi_1(\boldsymbol{x}), \phi_2(\boldsymbol{x}), \dots, \phi_n(\boldsymbol{x})\big)$ being the shape function and

$$\phi_i(\boldsymbol{x}) = \sum_{j=1}^{m} p_j(\boldsymbol{x})[\boldsymbol{A}^{-1}(\boldsymbol{x})\boldsymbol{B}(\boldsymbol{x})]_{ji}. \tag{16}$$

The spatial derivatives of the shape functions are computed as

$$\phi_{i,k}(\boldsymbol{x}) = \sum_{j=1}^{m} p_{j,k}(\boldsymbol{A}^{-1}\boldsymbol{B})_{ji} + p_j\big(\boldsymbol{A}_{,k}^{-1}\boldsymbol{B} + \boldsymbol{A}^{-1}\boldsymbol{B}_{,k}\big)_{ji} \tag{17}$$

with $\boldsymbol{A}_{,k}^{-1} = -\boldsymbol{A}^{-1}\boldsymbol{A}_{,k}\boldsymbol{A}^{-1}$ and the index following a comma denoting a spatial derivative.

<u>*Modified Moving Least Squares (MMLS)*</u>

The process of constructing shape functions and their derivatives (Eq. (16)-(17)) described above fails when the moment matrix $\boldsymbol{A}$ in Eq. (12) is singular. Singularity of the moment matrix is caused by certain degenerate distributions of nodes within the spatial domain. This results in some node distributions to be *inadmissible* [1, 10, 17]. To allow almost arbitrary node placement, i.e. to have almost all nodal distribution *admissible*, we suggested the following modification to Moving Least Square approximation [10].

As with MLS, we start with the approximation of a function $u(\boldsymbol{x})$, denoted by $u^h(\boldsymbol{x})$, which is defined by a combination of $m$ monomials (Eq. (7)). As in MLS, these coefficients are computed by minimizing an error functional which is defined based on the weighted least squares errors and including additional constraints on the coefficients $\boldsymbol{a}$ corresponding to the second-degree monomials in the basis. By adding constraints to the error functional we can ensure the invertibility of the moment matrix $\boldsymbol{A}(\boldsymbol{x})$ for all nodal distributions which are admissible when using first degree monomials as base functions in MLS. In this approach the classical MLS solution is altered only very slightly when the moment matrix is not singular.

The error functional is defined as

$$J(\mathbf{x}) = \sum_{j=1}^{n} \left[\big(u^h(\mathbf{x}_j) - u_j\big)^2 w\big(\|\mathbf{x} - \mathbf{x}_j\|\big)\right] + \mu_{x^2}a_{x^2}^2 + \mu_{y^2}a_{y^2}^2 + \mu_{z^2}a_{z^2}^2 + \mu_{xy}a_{xy}^2 + \mu_{xz}a_{xz}^2 + \mu_{yz}a_{yz}^2, \tag{18}$$

where $n$ is the number of nodes in the support of $\boldsymbol{x}$, and $\mu = [\mu_{x^2}\ \mu_{y^2}\ \mu_{z^2}\ \mu_{xy}\ \mu_{xz}\ \mu_{yz}]$ are positive (small) weights for the additional constraints. After minimization and solving the resulting systems of equations, the MMLS approximation is obtained as

$$u^h(\boldsymbol{x}) = \mathbf{p}^T(\mathbf{P}^T\mathbf{W}\mathbf{P} + \mathbf{H})^{-1}\mathbf{P}^T\mathbf{W}\boldsymbol{u} = \sum_{j=1}^{n} \Phi_j\,(\boldsymbol{x})u_j = \boldsymbol{\Phi}^{\boldsymbol{T}}(\mathbf{x})\boldsymbol{u}, \tag{19}$$

where $\boldsymbol{\Phi}$ is the vector of shape functions:

$$\boldsymbol{\Phi} = [\phi_1(\mathbf{x}) \ldots \phi_n(\mathbf{x})] = \mathbf{p}^T(\mathbf{P}^T\mathbf{W}\mathbf{P} + \mathbf{H})^{-1}\mathbf{P}^T\mathbf{W}. \tag{20}$$

In 2D, $\mathbf{H}$ is a $6 \times 6$ matrix with all elements equal to zero except the last three diagonal entries that are equal to the positive weights of the additional constraints $\mu$

$$\boldsymbol{H} = \begin{bmatrix} \mathbf{0}_{33} & \mathbf{0}_{33} \\ \mathbf{0}_{33} & \operatorname{diag}(\boldsymbol{\mu}) \end{bmatrix}. \tag{21}$$

In 3D, $\mathbf{H}$ is a $10 \times 10$ matrix with all elements equal to zero except the last six diagonal entries that are equal to $\mu$

$$\boldsymbol{H} = \begin{bmatrix} \mathbf{0}_{44} & \mathbf{0}_{46} \\ \mathbf{0}_{64} & \operatorname{diag}(\boldsymbol{\mu}) \end{bmatrix}. \tag{22}$$

*2.3 Regularized weight function and interpolating shape functions*

A very desirable property of shape functions is to be *interpolating* (rather than *approximating*). The choice of weight function regulates the interpolation properties of MMLS (Interpolating MMLS or IMMLS). In this study we use the regularized weight function proposed by Most and Bucher [19]

$$w_R(r) = \frac{\left(\left(\frac{r}{r_{SD}}\right)^2 + \epsilon\right)^{-2} - (1+\epsilon)^{-2}}{\epsilon^{-2} - (1+\epsilon)^{-2}}, \tag{23}$$

with $\epsilon \ll 1$ denoting a regularization parameter which must be very small such that the condition $w_i(\boldsymbol{x}_j) \approx \delta_{ij}$ is fulfilled with high accuracy. Fig. 1 displays the regularized weight function as a function of the normalized distance $r/r_{SD}$ and the positions of neighboring nodes ($r_{SD}$ is the radius of the support domain).

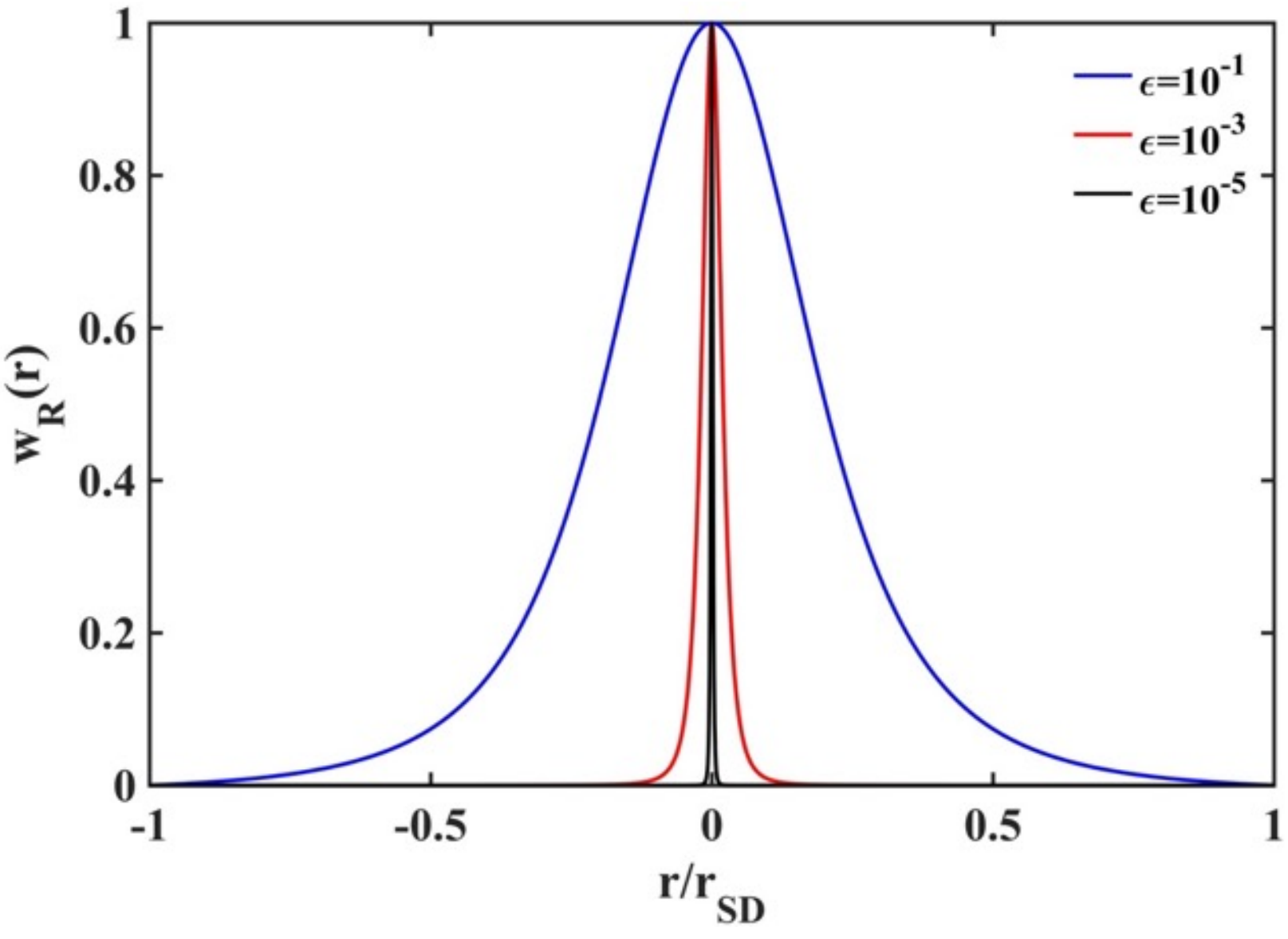

**Fig 1.** Weight function $w_R(r)$ for various values of the regularization parameter $\epsilon$ versus the normalized distance $r/r_{SD}$, where $r_{SD}$ is the radius of the support. As the regularization parameter decreases the weight function becomes steeper and approximates Kronecker delta.

Additionally, $\epsilon$ must be larger than the square root of machine precision to avoid numerical stability problems [18]. We set the value for the regularization parameter to $\epsilon = 10^{-5}$. The maximum difference between the regularized weight function and Kronecker delta function $(\delta_{ij})$ at the nodes in the support domain can be approximated as [19]

$$\left|w_i(\boldsymbol{x}_j) - \delta_{ij}\right|_{MAX} \approx \left(\left(\frac{r_{min}}{r_{SD}}\right)^{-4} - 1\right)\epsilon^2, \tag{24}$$

where $r_{min}$ is the minimum distance between two nodes in the support. After substituting the regularized weight function (Eq. 23) into a formula for the error functional (Eq. 24) and following derivation steps from the previous section we obtain shape function with the interpolation property for practical purposes

$$\Phi_i^{IMMLS}(\boldsymbol{x}_j) = \delta_{ij}. \tag{25}$$

As our $\Phi_i^{IMMLS}$ shape functions are interpolating, which is a consequence of our choice of the regularized weight function $w_R(r)$, Eq. (23), we can impose essential boundary conditions directly as in the finite element method.

In Figure 2 we show our $\Phi_i^{IMMLS}$ shape function. We also include MLS shape function with commonly applied exponential weight function [1, 17] for comparison. Figure 2 clearly demonstrates that by using the regularized weight function the interpolating property of the shape function is achieved (solid lines - interpolating moving least squares (IMMLS) shape function), in contrast to the shape function obtained with the exponential weight function (dashed lines).

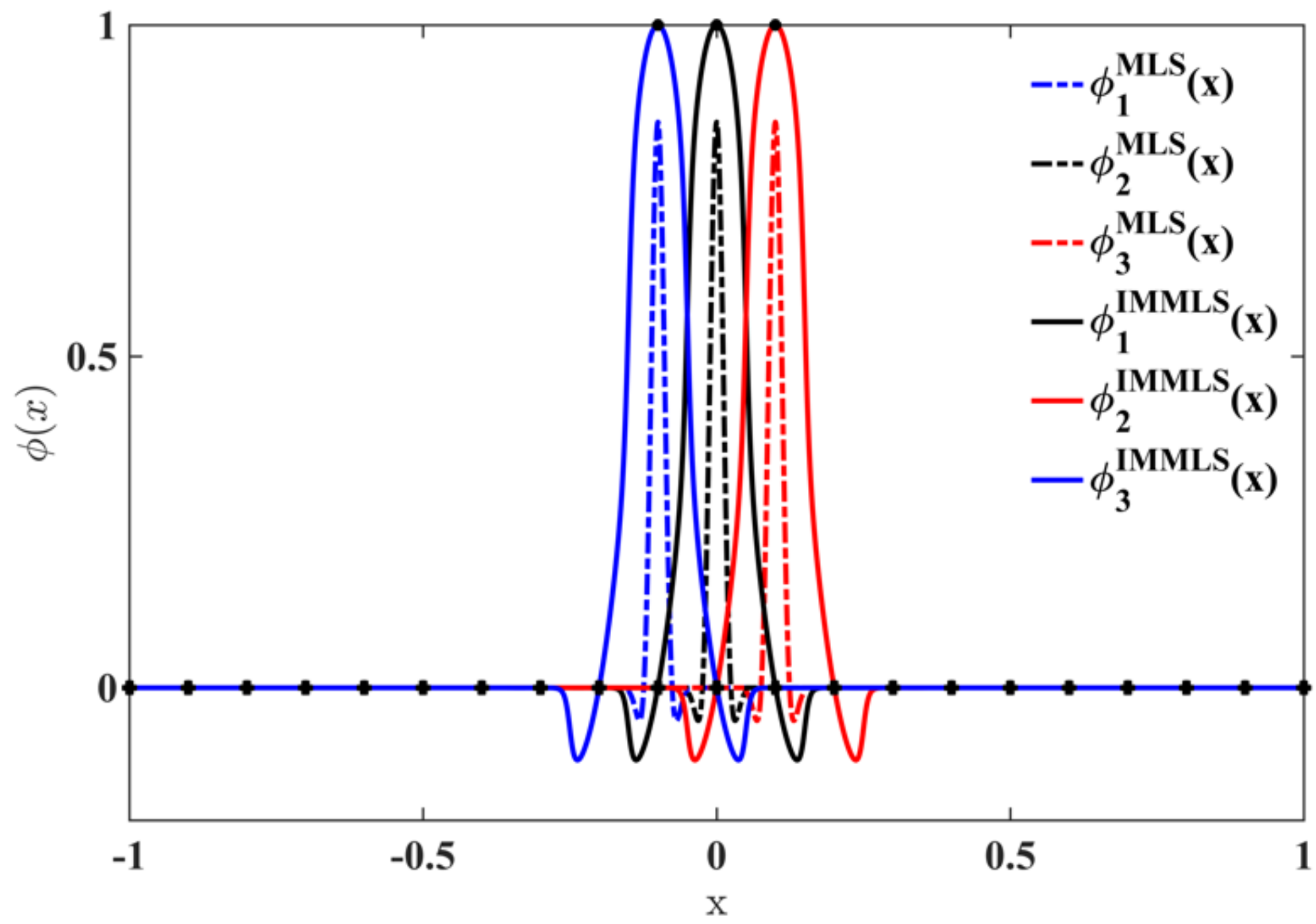

**Fig 2.** IMMLS shape functions with regularized weight function, compared to MLS shape functions with exponential weight function.

*2.4 Numerical integration*

Evaluating nodal reaction forces in Eq.(2) requires numerical integration over the volume of the problem domain. The integrands are not polynomials so that exact integration is very difficult and impossible to achieve using commonly employed numerical integration schemes such as Gaussian integration. Nevertheless, as shown in [11], extremely accurate volumetric integration is not necessary to obtain sufficiently accurate results in one's variable of interest. Therefore, in this work we use simple 4-point Gauss scheme over a domain conforming, automatically generated tetrahedral grid (that does not need to satisfy stringent quality criteria required by finite element method). In theory, this approach is not entirely satisfactory, but it gives very good results in practice, as shown in Section 3 and 4.

## 3. Algorithm verification

In this section we present 3D numerical examples which verify the accuracy of the presented methodology against analytical solutions (where available) and well-established finite element (FE) procedures (for moderate deformations, where FE analysis is expected to give reliable and accurate results). We consider four examples: ($i$) unconstrained compression of a cube; ($ii$) compression of a cylinder, ($iii$) extension of a cylinder, and ($iv$) shear of a cylinder. We pay special attention to demonstrating the accuracy of essential boundary condition imposition. Taking advantage of interpolating property of our regularized modified moving least square shape functions, we impose essential boundary conditions directly, as in the finite element method.

When comparing our meshless solutions to FEM, we used identical geometry, material models and material properties. In the FEM computations, we used a nonlinear static procedure based on a direct solver and Newton's iterative method available in ABAQUS [20]. In all examples, we used a hyperelastic neo-Hookean material model. This is the simplest model compatible with the finite deformation theory [21], and we have used it in many of our previous studies on computing soft continua and soft tissue deformations. The neo-Hookean material has the strain energy density functional $W = \frac{\mu}{2}(\bar{I}_1 - 3) + \frac{k}{2}(J - 1)^2$, where $\mu, k$ are the Lame parameters, $J$ and $\bar{I}_1 = \bar{\lambda}_1^2 + \bar{\lambda}_2^2 + \bar{\lambda}_3^2$ (with $\bar{\lambda}_i = J^{-1/3}\lambda_i$) are the determinant of the deformation gradient and the first invariant of the isochoric part of the right Cauchy-Green deformation tensor.

In all analyzed examples, the load was imposed through the motion of the boundary to demonstrate the simplicity and accuracy of boundary condition imposition. The displacement $\boldsymbol{u}(\boldsymbol{x})$ of a given node $\boldsymbol{x}$ is enforced over a period $T$ up to the maximum displacement $\boldsymbol{u}_{max}$ according to the smooth function (3-4-5 polynomial guaranteeing zero acceleration at the start and end of the loading phase)

$$\boldsymbol{u}(\boldsymbol{x}) = \boldsymbol{u}_{max}\left(10\left(\frac{t}{T}\right)^3 - 15\left(\frac{t}{T}\right)^4 + 6\left(\frac{t}{T}\right)^5\right), \quad 0 \leq t \leq T. \tag{26}$$

*3.1 Compression of a hyperelastic cube*

As the first example, we consider the unconstrained compression of a cube with edge length of 100 mm. The cube is compressed by displacing the top surface by 20 mm (i.e. 20% of the initial height), while it is constrained at the bottom face $(u_z = 0)$. The boundary conditions are as shown in Fig. 3. The cube's material is described using a hyperelastic neo-Hookean model with Young's modulus of $E = 3{,}000$ Pa, Poisson's ratio of $v = 0.49$ and density of $\rho = 1{,}000$ kg/m$^3$. For this example, the analytical solution for the displacement field is available even for finite deformations.

The analytical solution for uniaxial compression (the same applies for extension) [22] is obtained by considering the principal stretches in the axial ($\lambda_3 = \lambda$) and lateral ($\lambda_1 = \lambda_2 = \sqrt{\frac{J}{\lambda}}$) directions, where $\lambda$ is the axial stretch in the direction of loading and $J = \det {}_0^t\mathbf{X}$ (where ${}_0^t\mathbf{X}$ is the deformation gradient). The principal Cauchy stresses for a compressible neo-Hookean material with stored energy function $W = \frac{\mu}{2}(\bar{I}_1 - 3) + \frac{k}{2}(J-1)^2$ are then given by

$$\sigma_{11} = \sigma_{22} = \frac{\mu}{3J^{5/3}}\left(\frac{J}{\lambda} - \lambda^2\right) + \kappa(J-1) \tag{27}$$

$$\sigma_{33} = \frac{2\mu}{3J^{5/3}}\left(\lambda^2 - \frac{J}{\lambda}\right) + \kappa(J-1). \tag{28}$$

After taking the stress differences and considering the boundary conditions $\sigma_{11} = \sigma_{22} = 0$ we arrive at the following relation between the axial stretch and the volume ratio

$$\frac{\mu}{6}\left(\frac{J}{\lambda} - \lambda^2\right) + \frac{\kappa}{2}\left(J^{\frac{8}{3}} - J^{\frac{5}{3}}\right) = 0 \tag{29}$$

The displacement field for a specified axial stretch $\lambda$ can be obtained by solving equation Eq. (29) for the volume ratio $J$, and substituting into equations $\lambda_1 = \lambda_2 = \sqrt{\frac{J}{\lambda}}$ to obtain the stretch in the lateral directions.

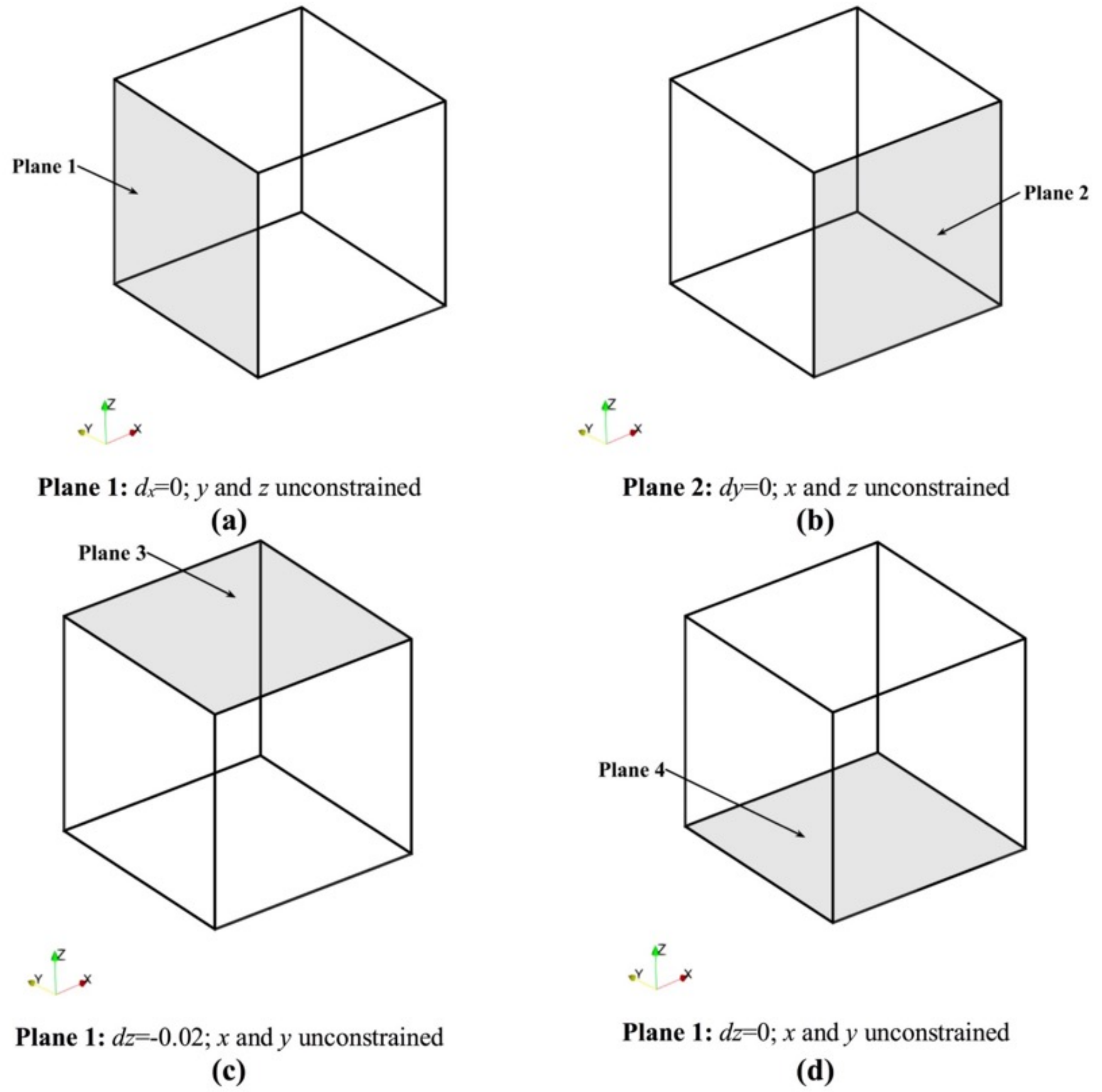

**Fig. 3.** Boundary conditions for modelling of unconstrained compression of a cube.

We use successively denser computational grids (point clouds) to ascertain a grid independent solution. The coarsest grid (Cloud 1) consists of $6 \times 6 \times 6$ (216 nodes corresponding to $h = 0.02$ m node spacing; 750 integration cells), and the densest (Cloud 4) of $41 \times 41 \times 41$ (68,921 nodes corresponding to $h = 0.0025$ m node spacing; 384,000 integration cells). Table 1 lists the grid configurations used in the simulations. We use the maximum absolute error $L_\infty = \left|u_{numerical} - u_{analytical}\right|$ and the normalized root mean square error $L_{NRMSE} = \frac{\frac{1}{N}\sqrt{\sum_{i=1}^{N}\left(u_i^{numerical} - u_i^{analytical}\right)}}{u_{max}^{analytical} - u_{min}^{analytical}}$ as measures of accuracy. Table 2 lists the $L_\infty$ and $L_{NRMSE}$ error norms of the proposed meshless method against the analytical solution, for the different nodal densities considered.

**Table 1.** Number of nodes, integration cells and grid spacing of the successively denser nodal distributions (clouds of points) for compression of hyperelastic cube.

| | **Cloud 1** | **Cloud 2** | **Cloud 3** | **Cloud 4** |
|---|---|---|---|---|
| **number of nodes** | 216 | 1,331 | 9,261 | 68,921 |
| **number of integration cells** | 750 | 6,000 | 48,000 | 384,000 |
| **spacing of points $h$ [m]** | 0.02 | 0.01 | 0.005 | 0.0025 |

**Table 2.** $L_\infty$ and $L_{NRMSE}$ error norm between the proposed meshless method and analytical solution for compression of hyperelastic cube.

| | Displacement component | $h = 0.02$ m | $h = 0.01$ m | $h = 0.005$ m | $h = 0.0025$ m |
|---|---|---|---|---|---|
| $L_\infty$ (in m) | $x$ | $5.81 \times 10^{-5}$ | $4.46 \times 10^{-5}$ | $1.43 \times 10^{-5}$ | $1.92 \times 10^{-5}$ |
| | $y$ | $8.71 \times 10^{-5}$ | $3.66 \times 10^{-5}$ | $1.57 \times 10^{-5}$ | $2.05 \times 10^{-5}$ |
| | $z$ | $5.69 \times 10^{-5}$ | $5.18 \times 10^{-5}$ | $3.50 \times 10^{-5}$ | $1.88 \times 10^{-5}$ |
| $L_{NRMSE}$ | $x$ | $9.44 \times 10^{-4}$ | $4.88 \times 10^{-4}$ | $2.29 \times 10^{-4}$ | $2.28 \times 10^{-4}$ |
| | $y$ | $1.07 \times 10^{-3}$ | $4.46 \times 10^{-4}$ | $2.33 \times 10^{-4}$ | $5.29 \times 10^{-4}$ |
| | $z$ | $6.48 \times 10^{-4}$ | $5.53 \times 10^{-4}$ | $2.64 \times 10^{-4}$ | $1.43 \times 10^{-4}$ |

As shown in Table 2, the numerical solution starts to converge and provide a grid independent solution for grid resolution better than $h = 0.01$ m (1,331 nodes).

To demonstrate the accuracy and the direct imposition of the essential boundary conditions, we compute the difference between the fictitious values $\boldsymbol{u} = (u_x, u_y, u_z)$ (Eq.(15)) and the approximated values $\boldsymbol{u}^h = (u_x^h, u_y^h, u_z^h)$ (Eq.(15)) of the displacement field, on the boundary nodes where the essential boundary conditions are prescribed. The size of this difference is the key reason why in many meshless methods direct imposition of essential boundary conditions is very inaccurate and elaborate treatments are necessary. Table 3 lists the maximum absolute difference $L_\infty^{(\boldsymbol{u},\boldsymbol{u}^h)} = |\boldsymbol{u} - \boldsymbol{u}^h|$ and the root mean square difference $L_2^{(\boldsymbol{u},\boldsymbol{u}^h)} = \frac{1}{N}\sqrt{\sum_{i=1}^{N}(\boldsymbol{u}_i - \boldsymbol{u}_i^h)}$ ($N$is the number of nodes) between $\boldsymbol{u}$ and $\boldsymbol{u}^h$**,** computed at the boundary nodes where essential boundary conditions are prescribed (as shown in Fig. 3). The error values listed in Table 3 were computed for the densest grid (Cloud 4 – 68,921 nodes and 384,000 integration cells).

**Table 3.** Compression of hyperelastic cube: maximum absolute difference $L_\infty^{(\boldsymbol{u},\boldsymbol{u}^h)}$ and root mean square error $L_2^{(\boldsymbol{u},\boldsymbol{u}^h)}$ between the fictitious ($\boldsymbol{u}$) and approximated ($\boldsymbol{u}^h$) displacement values at the boundary nodes where essential boundary condition are prescribed (Plane 1 - $u_x$, Plane 2 - $u_y$, Plane 3 and Plane 4 - $u_z$).

| | **Plane 1** | **Plane 2** | **Plane 3** | **Plane 4** |
|---|---|---|---|---|
| $L_\infty^{(\boldsymbol{u},\boldsymbol{u}^h)}$ (in m) | $5.69 \times 10^{-13}$ | $9.60 \times 10^{-13}$ | $1.12 \times 10^{-13}$ | $4.41 \times 10^{-12}$ |
| $L_2^{(\boldsymbol{u},\boldsymbol{u}^h)}$ (in m) | $7.06 \times 10^{-14}$ | $7.51 \times 10^{-14}$ | $3.44 \times 10^{-14}$ | $2.12 \times 10^{-13}$ |

Values in Table 3 demonstrate that direct imposition of essential boundary conditions is, for practical purposes, exact.

Fig. 4 shows the initial and the final configuration of the cube (results in Fig. 4 produced using 9,261 nodes, where the solution is already grid independent).

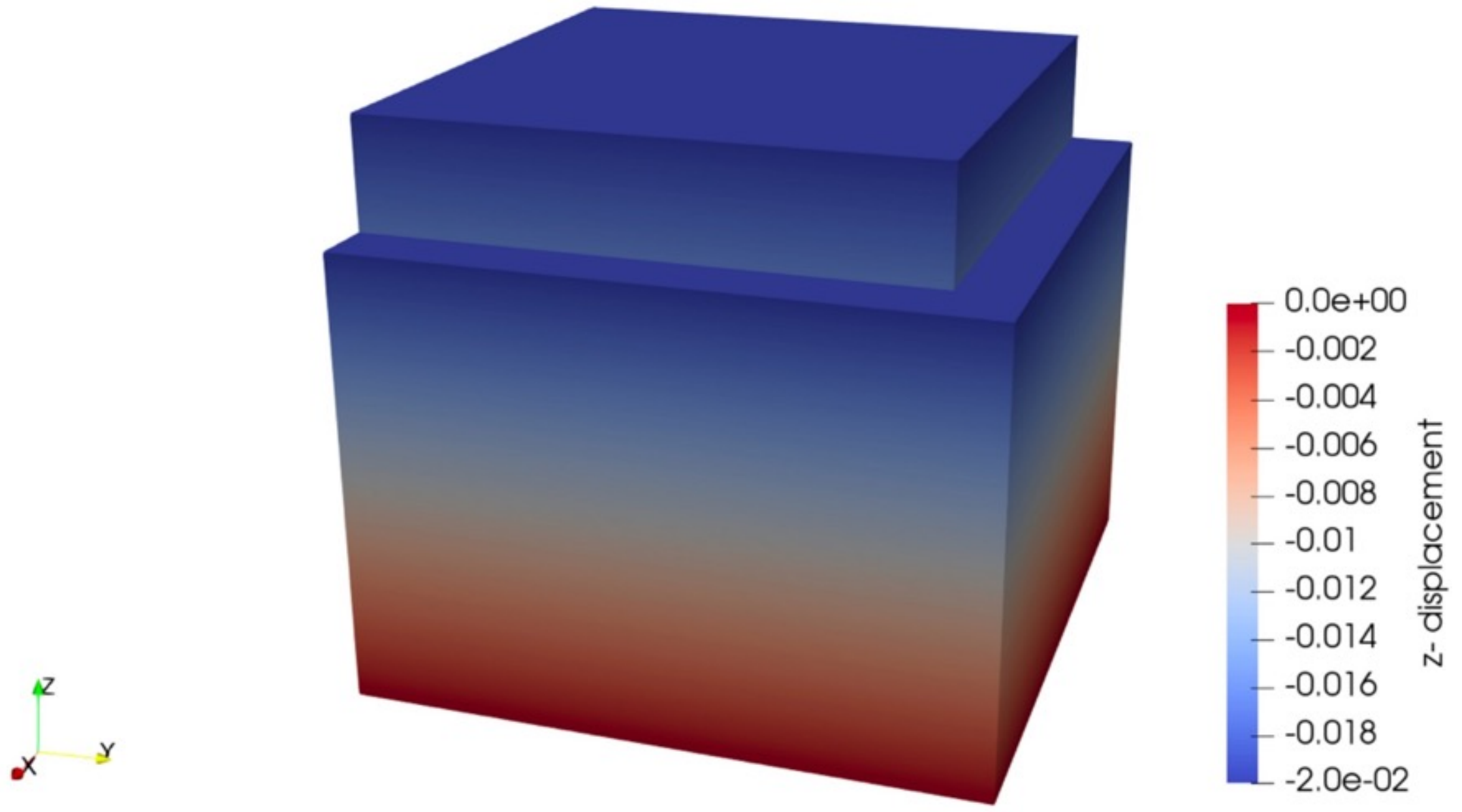

**Fig. 4.** Initial and the final configuration of the cube. The displacement field in the z- direction is shown; boundary conditions on the top and bottom surface are accurately imposed.

*3.2 Compression of a cylinder*

In this example, we consider the compression of a cylinder with height and diameter of 100 mm. The cylinder's material is described using a hyperelastic neo-Hookean model with Young's modulus of $E = 3{,}000$ Pa, Poisson's ratio of $v = 0.49$ and density of $\rho = 1{,}000$ kg/m$^3$. The cylinder is compressed by 20 mm by displacing the top surface along the vertical direction ($u_x = u_y = 0$), while it is rigidly constrained at the bottom face $\left(u_x = u_y = u_z = 0\right)$.

We use successively denser computational grid configurations to obtain a grid independent numerical solution. When discretizing the analyzed spatial domain, we refine the coarse grid by adding nodes to create denser grids, i.e. the moderate and denser grids inherit the nodes from the coarse grid. This way, it is straightforward to compare solutions for nodal variables, such as displacements. Table 4 lists the grid configurations used in the simulations.

**Table 4.** Number of nodes, integration cells and average grid spacing of the successively denser nodal distributions (clouds of nodes) for the compression of a cylinder verification example.

| | Cloud 1 | Cloud 2 | Cloud 3 |
|---|---|---|---|
| **number of nodes** | 1,089 | 7,769 | 58,641 |
| **number of integration cells** | 5,184 | 41,472 | 331,776 |
| **average spacing of points *h* [mm]** | 8.22 | 4.44 | 2.65 |

For cylinder compression, an analytical solution for the vertical component of the displacement field is available at the plane of symmetry of the cylinder $z = 50$ mm ( $u_z = 0.5u_z^{max}$), with $u_z^{max} = -20$ mm applied at the top face. The analytical solution for the entire displacement field was reported in [23]. However, it was later shown to be inaccurate because of incorrect simplifying assumption used in derivation [24]. Table 5 lists the $L_\infty$ and $L_{NRMSE}$ between our meshless solution and the analytical solution.

**Table 5.** Compression of a cylinder example. $L_\infty$ and $L_{NRMSE}$ error norm between the proposed meshless method and analytical solution for horizontal plane $z = 50$ mm.

| | **Cloud 1** | **Cloud 2** | **Cloud 3** |
|---|---|---|---|
| $L_\infty$ (in m) | $1.49 \times 10^{-4}$ | $5.34 \times 10^{-5}$ | $3.62 \times 10^{-5}$ |
| $L_{NRMSE}$ | $1.17 \times 10^{-3}$ | $4.23 \times 10^{-4}$ | $2.85 \times 10^{-4}$ |

From Table 5, we can observe that the solution has converged for 7,769 nodes, such that higher number of nodes provide a grid independent numerical solution. Fig. 5a shows the axonometric view of the cylinder compressed by 20% of its initial height (compression by 20 mm), while Fig. 5b shows a cross section of the compressed cylinder in X-Z plane at $y = 0$.

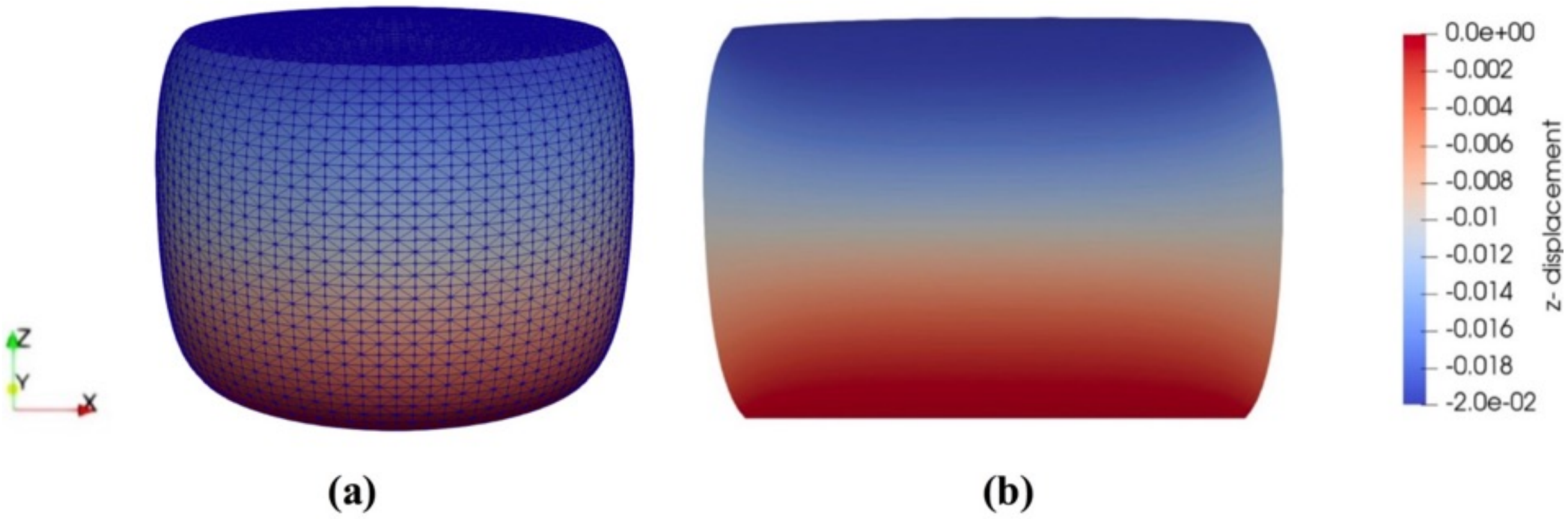

**Fig. 5. (a)** Final configuration of the cylinder under 20% compression. The displacement field in the z- direction is shown; boundary conditions on the top and bottom surface are accurately imposed and **(b)** cross-section of the compressed cylinder in X-Z plane at $y = 0$.

To demonstrate the accuracy of the direct imposition of the essential boundary conditions, we compute the difference between the fictitious values of the displacement field $\boldsymbol{u} = (u_x, u_y, u_z)$ (Eq.8) and the approximated values $\boldsymbol{u}^h = (u_x^h, u_y^h, u_z^h)$ (Eq.15) on the boundary nodes where the essential boundary conditions are prescribed. Table 6 lists the $L_\infty^{(\boldsymbol{u},\boldsymbol{u}^h)}$ and the $L_2^{(\boldsymbol{u},\boldsymbol{u}^h)}$error norms between $\boldsymbol{u}$ and $\boldsymbol{u}^h$, computed at the bottom ($u_x = u_y = u_z = 0$) and top ($u_x = u_y = 0, u_z = -0.02$ m) surface. The $L_\infty^{(\boldsymbol{u},\boldsymbol{u}^h)}$ and the $L_2^{(\boldsymbol{u},\boldsymbol{u}^h)}$ error norms listed in Table 6 were computed on Cloud 3 (58,641 nodes and 331,776 integration cells). As in the previous example of a compressed cube, our method allows, for practical purposes, exact application of essential boundary conditions.

**Table 6.** Maximum absolute difference $L_\infty^{(\boldsymbol{u},\boldsymbol{u}^h)}$ and root mean square error $L_2^{(\boldsymbol{u},\boldsymbol{u}^h)}$ between the fictitious $(\boldsymbol{u})$ and approximated $(\boldsymbol{u}^h)$ at the boundary nodes located at the bottom $(z = 0)$ and top $(z = 0.1$ m) surface of the compressed cylinder.

| | **Displacement component** | **Bottom ($\boldsymbol{z = 0}$)** | **Top ($\boldsymbol{z = 0.1}$ m)** |
|---|---|---|---|
| $L_\infty^{(\boldsymbol{u},\boldsymbol{u}^h)}$ (in m) | $x$ | $8.12 \times 10^{-11}$ | $8.49 \times 10^{-11}$ |
| | $y$ | $8.10 \times 10^{-11}$ | $8.50 \times 10^{-11}$ |
| | $z$ | $4.38 \times 10^{-11}$ | $1.14 \times 10^{-13}$ |
| $L_2^{(\boldsymbol{u},\boldsymbol{u}^h)}$ (in m) | $x$ | $1.21 \times 10^{-11}$ | $1.20 \times 10^{-11}$ |
| | $y$ | $1.21 \times 10^{-11}$ | $1.20 \times 10^{-11}$ |
| | $z$ | $8.43 \times 10^{-12}$ | $4.30 \times 10^{-14}$ |

### *3.3 Extension of a cylinder*

In this example, we consider the extension of the cylinder described in the previous section. The cylinder is extended by 100 mm (by displacing the top surface by 100 mm – 100% along the vertical axis, $(u_x = u_y = 0)$ ), while it is rigidly constrained at the bottom face $(u_x = u_y = u_z = 0)$.

We use a dense grid of 12,957 nodes and 39,633 integration cells, which based on the convergence analysis presented in the case of cylinder compression, ensures a grid independent numerical solution. As in the previous case, in the horizontal plane of symmetry of the cylinder the analytical solution for the vertical component of the displacement field is available. We demonstrate the accuracy of our methods by comparing the numerical solution with the analytical solution. Table 7 lists the $L_\infty$ and $L_{NRMSE}$ between our numerical solution and analytical solution. Fig. 6 shows the final configuration of the cylinder under 100% extension.

**Table 7.** $L_\infty$ and $L_{NRMSE}$ error norm between the proposed meshless method and analytical solution for cylinder extension.

| | $L_\infty$ (in m) | $L_{NRMSE}$ |
|---|---|---|
| 12,957 nodes and 39,633 integration cells | $1.49 \times 10^{-4}$ | $1.17 \times 10^{-3}$ |

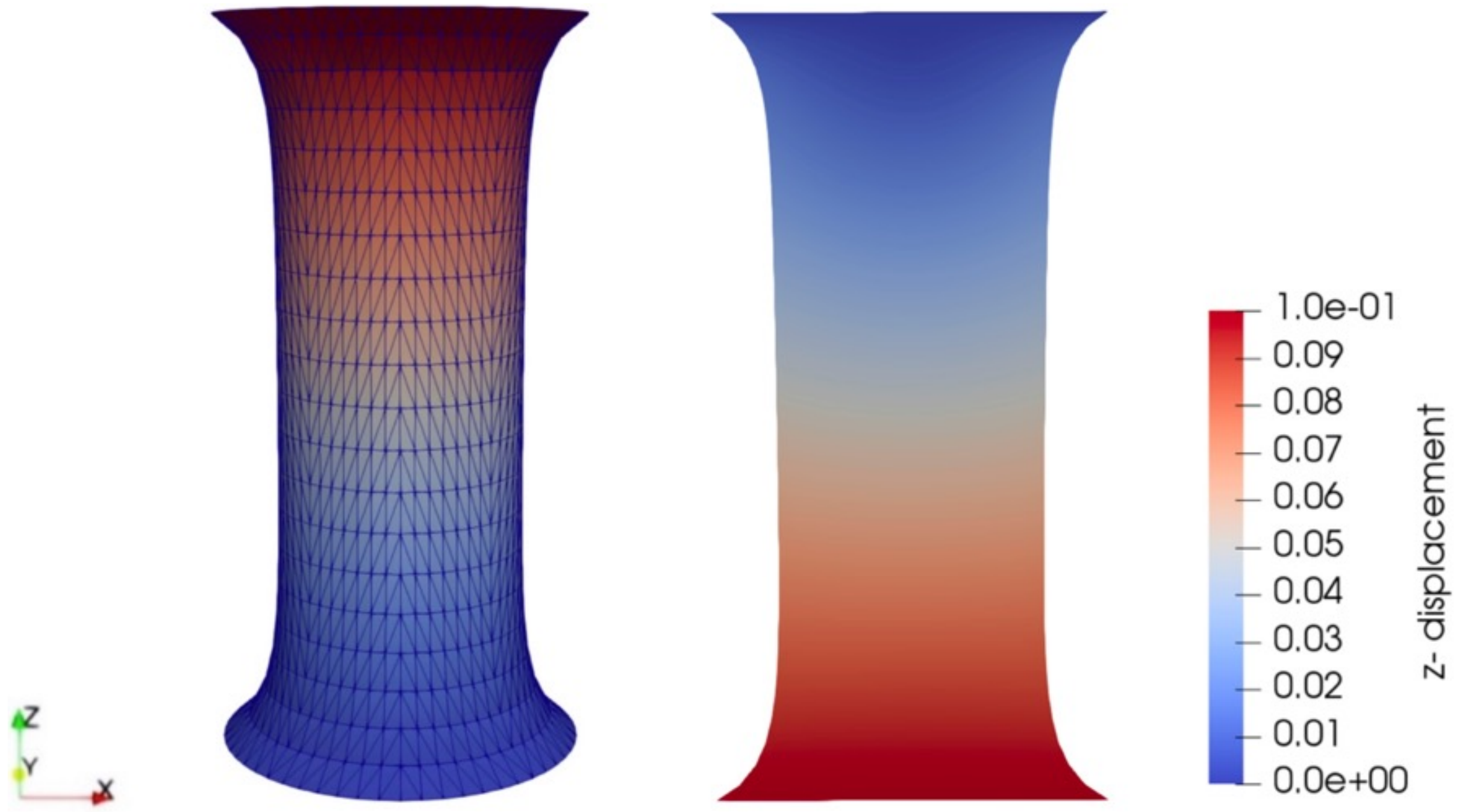

**(a)** **(b)**

**Fig. 6.** **(a)** Final configuration of the cylinder under 100% extension. The displacement field in the *z*- direction is shown; boundary conditions on the top and bottom surface are accurately imposed and **(b)** cross-section at $y = 0$.

We demonstrate the accuracy of direct imposition of the essential boundary conditions by computing the difference between the fictitious values of the displacement field $\boldsymbol{u}$ and the approximated values $\boldsymbol{u}^h$ on the bottom and top surfaces where the essential boundary conditions are prescribed. Table 8 lists the $L_\infty^{(\boldsymbol{u},\boldsymbol{u}^h)}$ and the $L_2^{(\boldsymbol{u},\boldsymbol{u}^h)}$error norms between $\boldsymbol{u}$ and $\boldsymbol{u}^h$, computed at the bottom ($u_x = u_y = u_z = 0$) and top ($u_x = u_y = 0, u_z = 0.1$ m) surface.

**Table 8.** Maximum absolute difference $L_\infty^{(\boldsymbol{u},\boldsymbol{u}^h)}$ and root mean square error $L_2^{(\boldsymbol{u},\boldsymbol{u}^h)}$ between the fictitious ($\boldsymbol{u}$) and approximated ($\boldsymbol{u}^h$) displacement field at the boundary nodes located at the bottom ($z = 0$) and top ($z = 0.1$ m) surface of the cylinder subjected to extension.

| | **Displacement component** | **Bottom ($z = 0$)** | **Top ($z = 0.1$)** |
|---|---|---|---|
| $L_\infty^{(\boldsymbol{u},\boldsymbol{u}^h)}$ (in m) | $x$ | $2.84 \times 10^{-10}$ | $2.75 \times 10^{-10}$ |
| | $y$ | $2.84 \times 10^{-10}$ | $2.75 \times 10^{-10}$ |
| | $z$ | $5.26 \times 10^{-11}$ | $8.24 \times 10^{-11}$ |
| $L_2^{(\boldsymbol{u},\boldsymbol{u}^h)}$ (in m) | $x$ | $5.87 \times 10^{-11}$ | $5.60 \times 10^{-11}$ |
| | $y$ | $5.87 \times 10^{-11}$ | $5.60 \times 10^{-11}$ |
| | $z$ | $1.36 \times 10^{-11}$ | $1.31 \times 10^{-11}$ |

### 3.4 Shear of a cylinder

We consider the shear of the same cylinder as in the previous sections. The cylinder's top surface is displaced in the horizontal $x$- direction by 50 mm ($u_y = u_z = 0$), while it is rigidly constrained at the bottom face $(u_x = u_y = u_z = 0)$.

As before, to assure grid-independent solution, we use a dense grid consisting of 12,957 nodes and 39,633 integration cells. We demonstrate the accuracy of the proposed scheme by comparing the numerical solution computed using the proposed meshless method, with the numerical results computed using ABAQUS non-linear finite element code [20]. Table 9 lists the $L_\infty$ and $L_{NRMSE}$ between the displacement field computed using the proposed meshless numerical solution and the ABAQUS solutions obtained using two types of elements: quadratic (10-noded) C3D10 (98769 nodes and 71,736 elements) and hexahedral (20-noded) C3D20 (50,581 nodes 12,348 elements). For the purpose of calculating the error norms we use the numerical solutions obtained using ABAQUS as the reference.

We can observe that the relative difference in computed displacements between the proposed meshless method and ABAQUS (C3D10 and C3D20 element types) is between 1.32% and 2.17% (accounting for all displacement components). In the absence of the analytical solution it is difficult to judge which of the three solutions is most accurate. Fig. 7a shows the shear of the cylindrical sample up to 50 mm, produced using 12,957 nodes and 71,736 elements. Fig. 7b show a cross section of the cylindrical sample at $y = 0$.

**Table 9.** $L_\infty$ and $L_{NRMSE}$ error norm between the MTLED method and ABAQUS nonlinear finite element code (quadratic tetrahedral (10-noded) C3D10 and hexahedral (20-noded) C3D20 elements) for cylinder in shear.

| | **Displacement component** | **MTLED vs ABAQUS (C3D10)** | **MTLED vs ABAQUS (C3D20)** |
|---|---|---|---|
| $L_\infty$ (in m) | $x$ | $1.28 \times 10^{-3}$ | $1.22 \times 10^{-3}$ |
| | $y$ | $7.16 \times 10^{-4}$ | $6.29 \times 10^{-4}$ |
| | $z$ | $2.22 \times 10^{-3}$ | $2.19 \times 10^{-3}$ |
| $L_{NRMSE}$ | $x$ | $1.32 \times 10^{-2}$ | $1.39 \times 10^{-2}$ |
| | $y$ | $1.86 \times 10^{-2}$ | $2.03 \times 10^{-2}$ |
| | $z$ | $2.02 \times 10^{-2}$ | $2.17 \times 10^{-2}$ |

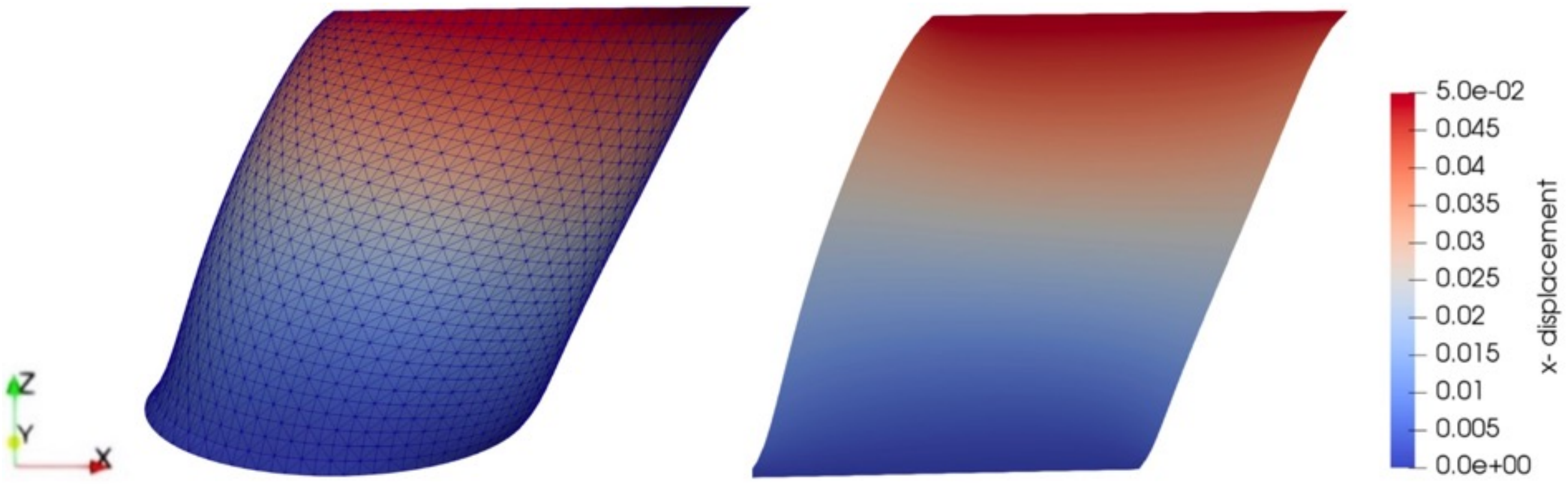

**(a)** **(b)**

**Fig. 7. (a)** Final configuration of the cylinder under 50% shear. The displacement field in the $x$- direction is shown; boundary conditions on the top and bottom surface are accurately imposed and **(b)** cross-section at $y = 0$.

As before, we demonstrate the accuracy and the direct imposition of the essential boundary conditions, by computing the difference between the fictitious values of the displacement field $\boldsymbol{u}$ and the approximated values $\boldsymbol{u}^h$ on the bottom and top surfaces where the essential boundary conditions are prescribed. Table 10 lists the $L_\infty^{(\boldsymbol{u},\boldsymbol{u}^h)}$ and the $L_2^{(\boldsymbol{u},\boldsymbol{u}^h)}$error norms between $\boldsymbol{u}$ and $\boldsymbol{u}^h$, computed at the bottom ($u_x = u_y = u_z = 0$) and top ($u_x = u_y = 0, u_z = 0.1$ m) surface.

**Table 10.** Maximum absolute difference $L_\infty^{(\boldsymbol{u},\boldsymbol{u}^h)}$ and root mean square error $L_2^{(\boldsymbol{u},\boldsymbol{u}^h)}$ between the fictitious ($\boldsymbol{u}$) and approximated ($\boldsymbol{u}^h$) at the boundary nodes located at the bottom ($z = 0$) and top ($z = 0.1$ m) surface of the cylinder in shear.

| | **Displacement component** | **Bottom ($z = 0$)** | **Top ($z = 0.1$ m)** |
|---|---|---|---|
| $L_\infty^{(\boldsymbol{u},\boldsymbol{u}^h)}$ (in m) | $x$ | $1.60 \times 10^{-10}$ | $1.01 \times 10^{-10}$ |
| | $y$ | $1.05 \times 10^{-10}$ | $1.13 \times 10^{-10}$ |
| | $z$ | $9.78 \times 10^{-11}$ | $9.80 \times 10^{-11}$ |
| $L_2^{(\boldsymbol{u},\boldsymbol{u}^h)}$ (in m) | $x$ | $2.79 \times 10^{-11}$ | $1.01 \times 10^{-10}$ |
| | $y$ | $2.13 \times 10^{-11}$ | $2.35 \times 10^{-11}$ |
| | $z$ | $1.55 \times 10^{-11}$ | $1.56 \times 10^{-11}$ |

## 4. Numerical examples

In this section, we further evaluate the accuracy of our meshless approach on challenging examples of cube twisting and very deep cylinder indentation, demonstrating that we can compute robust results well beyond what the finite element method can offer. Additionally, we demonstrate the applicability of our methods in the analysis of continua with complex geometry.

### *4.1 Torsion of a hyperelastic cube*

We model torsion of a cube made from a hyperelastic material. The cube has an edge length of 1 m and is described using the neo-Hookean material model with initial Young's modulus of $E = 3{,}000$ Pa, Poisson's ratio of $v = 0.3$ and density of $\rho = 1{,}000$ kg/m$^3$. The displacement boundary conditions that impose the cube twisting by $30^{\circ}$ are defined on Dirichlet boundary

$$\boldsymbol{u}|_{x=0} = \begin{pmatrix} 0 \\ 0 \\ 0 \end{pmatrix} \tag{28}$$

$$\boldsymbol{u}|_{x=0.01} = \begin{pmatrix} 0 \\ 0.5\left(0.5 + (y-0.5)cos\left(\frac{\pi}{2}-\frac{\pi}{6}\right) - (z-0.5)sin\left(\frac{\pi}{2}-\frac{\pi}{6}\right) - y\right) \\ 0.5\left(0.5 + (y-0.5)sin\left(\frac{\pi}{2}-\frac{\pi}{6}\right) + (z-0.5)cos\left(\frac{\pi}{2}-\frac{\pi}{6}\right) - z\right) \end{pmatrix}. \tag{29}$$

To ensure a grid-independent solution, we use a dense grid of 17,576 nodes and 93,750 integration cells. We demonstrate the accuracy of the proposed scheme by comparing the numerical solution computed using our meshless method with the numerical results computed using FEniCS [25]. FEniCS is a popular open-source computing platform for solving partial differential equations (PDEs), by translating scientific models into efficient finite element code. FEniCS allows straightforward definition of the boundary motion defined by Eq. (29), whereas in ABAQUS non-linear finite element code it is very difficult to do. Table 11 lists the $L_{\infty}$ and $L_{NRMSE}$ between the meshless and FEniCS solution [26], computed using 17,756 nodes and 93,750 linear tetrahedral elements.

**Table 11.** $L_\infty$ and $L_{NRMSE}$ error norm between the proposed meshless method and FEniCS for the twisting of a cube test case solved using 17,576 nodes and 93,750 integration cells.

| ***Displacement component*** | $\boldsymbol{L_\infty}$ **(in m)** | $\boldsymbol{L_{NRMSE}}$ |
|---|---|---|
| $x$ | $7.81 \times 10^{-3}$ | $1.65 \times 10^{-3}$ |
| $y$ | $4.87 \times 10^{-3}$ | $8.29 \times 10^{-4}$ |
| $z$ | $5.23 \times 10^{-3}$ | $8.55 \times 10^{-4}$ |

To highlight the robustness of our meshless scheme, we considered the case of twisting the cube with Poisson's ratio of $v = 0.49$, where FEniCS failed to provide results (even with quadratic 10-noded tetrahedral elements), most likely due to the severe distortion of the elements. Fig. 8 shows the deformation of the cube applying twisting by $30^{\circ}$.

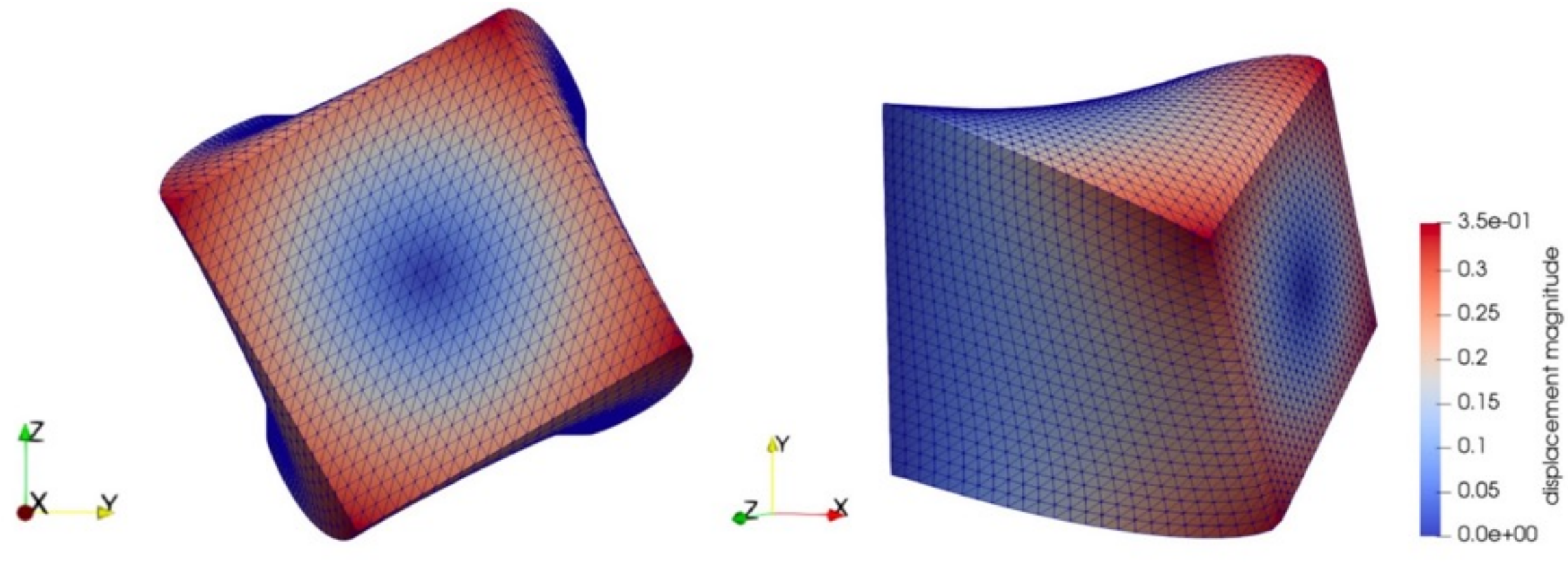

**(a)**

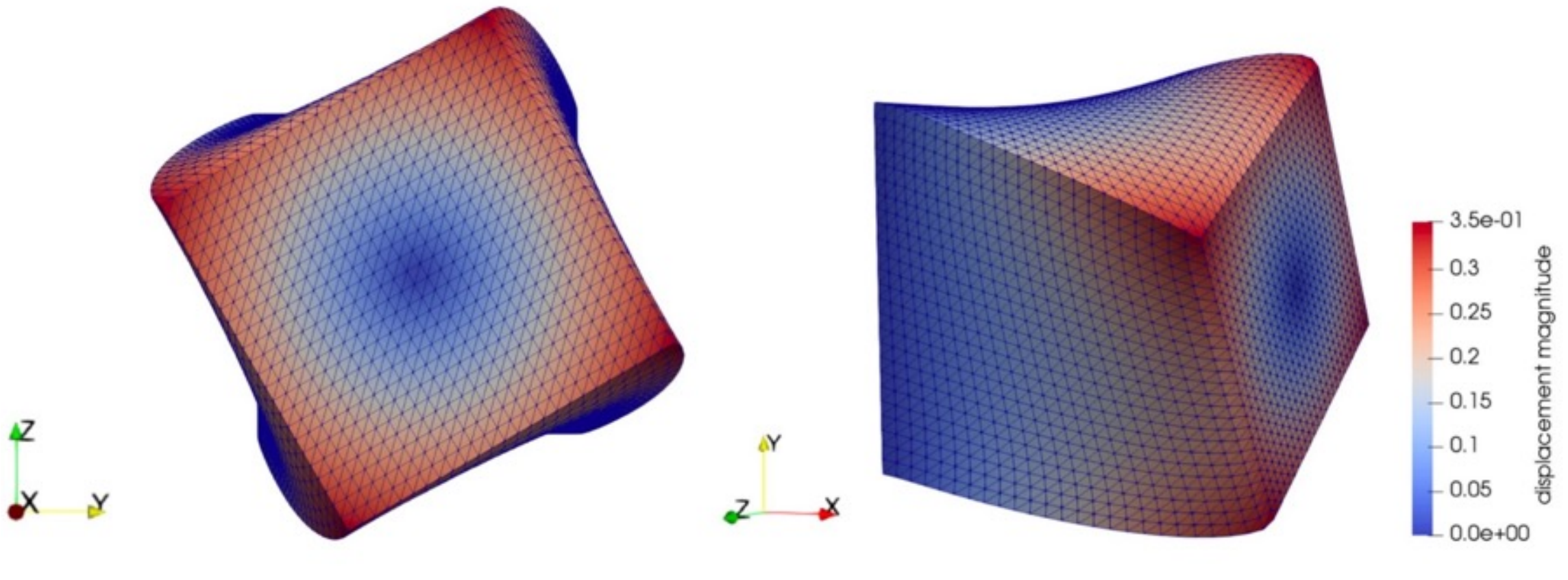

**(b)**

**Fig. 8.** Deformation of a neo-Hookean cube of dimensions $x = y = z = 1$ m after twisting ~~it~~ by $60^{\circ}$ for **(a)** $E = 3{,}000$ and $v = 0.3$ **(b)** $E = 3{,}000$ and $v = 0.49$.

As in the previous examples, to demonstrate the accuracy of essential boundary condition imposition, in Table 12 we list the $L_\infty^{(\boldsymbol{u},\boldsymbol{u}^h)}$ and the $L_2^{(\boldsymbol{u},\boldsymbol{u}^h)}$ error norms, computed at the $x = 0$ and $x = 1$ m surface of the cube.

**Table 12.** Maximum absolute difference $L_\infty^{(\boldsymbol{u},\boldsymbol{u}^h)}$ and root mean square error $L_2^{(\boldsymbol{u},\boldsymbol{u}^h)}$ between the fictitious $(\boldsymbol{u})$ and approximated $(\boldsymbol{u}^h)$ at the boundary nodes located at $x = 0$ and $x = 1$ m of the cube subjected to twisting as shown in Fig. 8.

| | **Displacement component** | $\boldsymbol{x = 0}$ | $\boldsymbol{x = 1}$ **m** |
|---|---|---|---|
| $L_\infty^{(\boldsymbol{u},\boldsymbol{u}^h)}$ (in m) | $x$ | $7.99 \times 10^{-11}$ | $6.53 \times 10^{-11}$ |
| | $y$ | $3.68 \times 10^{-11}$ | $6.23 \times 10^{-11}$ |
| | $z$ | $5.37 \times 10^{-11}$ | $5.92 \times 10^{-11}$ |
| $L_2^{(\boldsymbol{u},\boldsymbol{u}^h)}$ (in m) | $x$ | $1.13 \times 10^{-11}$ | $1.14 \times 10^{-11}$ |
| | $y$ | $6.89 \times 10^{-12}$ | $7.29 \times 10^{-12}$ |
| | $z$ | $8.43 \times 10^{-12}$ | $9.33 \times 10^{-12}$ |

### *4.2 Cylinder indentation*

In this example, we demonstrate the convergence of the proposed scheme for indentation of a cylinder with height of 17 mm and diameter of 30 mm (Fig. 9). As before, we take the material to be neo-Hookean, with Young's modulus of $E = 3{,}000$ Pa, Poisson's ratio of $\nu = 0.49$, and density of $\rho = 1{,}000$ kg/m$^3$. The cylinder is indented by 4 mm on the top surface ($u_x = u_y = 0$), while it is rigidly constrained at the bottom face $(u_x = u_y = u_z = 0)$ (Fig. 9).

To demonstrate convergence, we used successively denser point clouds, as shown in Table 13.

**Table 13.** Number of nodes and integration cells of the successively denser nodal distributions used in cylinder indentation example.

| | **Cloud 1** | **Cloud 2** | **Cloud 3** | **Cloud 4** |
|---|---|---|---|---|
| **number of nodes** | 2,581 | 7,291 | 31,860 | 51,035 |
| **number of integration cells** | 13,320 | 37,954 | 206,200 | 280,146 |

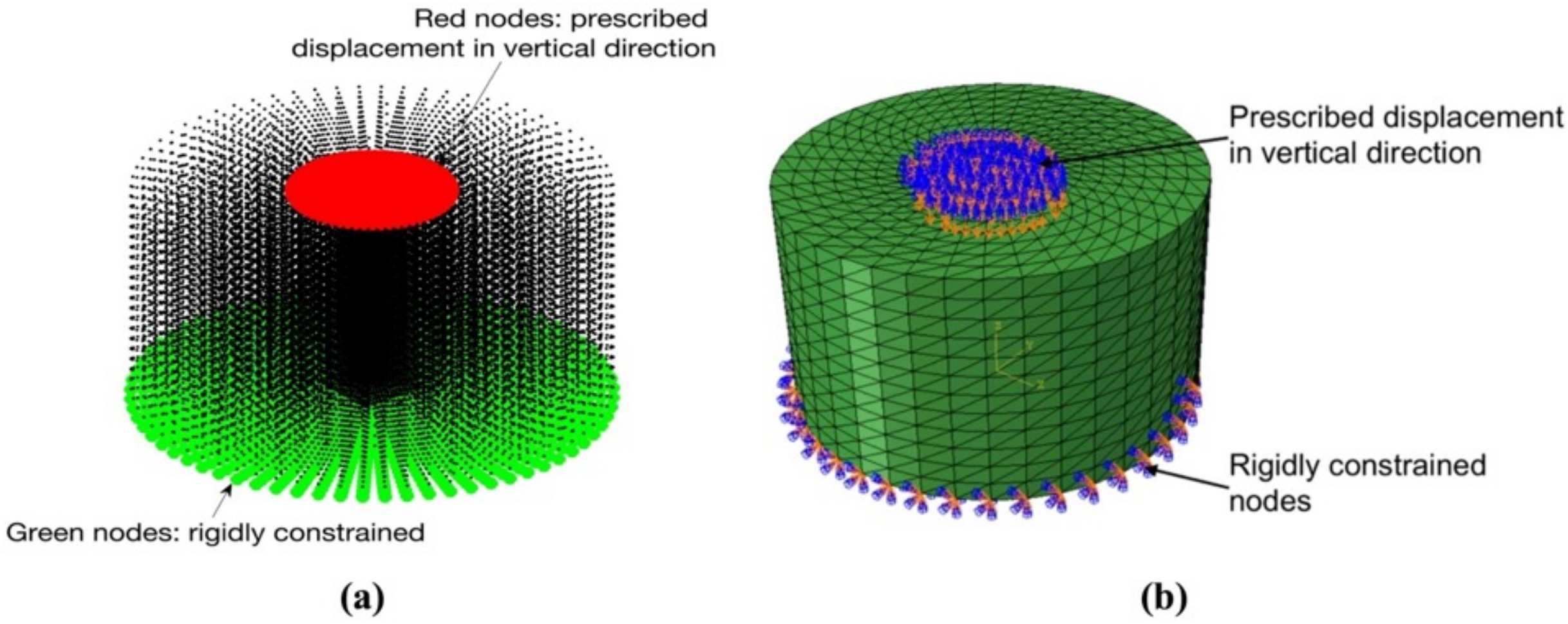

**Fig. 9.** Simulation of indentation of a cylinder **(a)** meshless model consisting of 31,860 nodes and 824,800 integration points in undeformed configuration. The prescribed displacement was applied at the nodes shown in red color. The nodes shown in green were rigidly constrained and **(b)** finite element model (used in ABAQUS non-linear code) consisting of 206,200 10-noded tetrahedral elements (elements C3D10 in ABAQUS) in undeformed configuration. The boundary conditions are the same as for the meshless model.

Table 14 lists the $L_\infty$ and $L_{NRMSE}$ for the numerical solution computed using the successively denser grids. The results obtained with the finest grid are taken as reference. The accuracy of the proposed meshless scheme increases with increasing number of nodes and already Cloud 3 offers a converged solution (see Table 14).

**Table 14.** $L_\infty$ and $L_{NRMSE}$ error norm for the successively denser nodal distributions (clouds of points) for cylinder indentation example.

| | **Displacement component** | **Cloud 1 to Cloud 4** | **Cloud 2 to Cloud 4** | **Cloud 3 to Cloud 4** |
|---|---|---|---|---|
| $L_\infty$ (in m) | $x$ | $4.64 \times 10^{-4}$ | $3.15 \times 10^{-4}$ | $2.06 \times 10^{-4}$ |
| | $y$ | $4.17 \times 10^{-4}$ | $2.47 \times 10^{-4}$ | $2.14 \times 10^{-4}$ |
| | $z$ | $5.79 \times 10^{-4}$ | $3.62 \times 10^{-4}$ | $2.51 \times 10^{-4}$ |
| $L_{NRMSE}$ | $x$ | $4.70 \times 10^{-2}$ | $2.20 \times 10^{-2}$ | $9.13 \times 10^{-3}$ |
| | $y$ | $4.26 \times 10^{-2}$ | $2.05 \times 10^{-2}$ | $9.12 \times 10^{-3}$ |
| | $z$ | $4.32 \times 10^{-2}$ | $2.30 \times 10^{-2}$ | $9.06 \times 10^{-3}$ |

We also compared the displacement field predicted using the proposed meshless method with that computed using established nonlinear algorithms available in ABAQUS (version 6.14) finite element code. In ABAQUS, we use 206,200 quadratic tetrahedral C3D10 elements.

**Table 15.** $L_\infty$ and $L_{NRMSE}$ error norm between the proposed meshless method and ABAQUS non-linear finite element code for cylinder indentation by 4 mm (24% of initial height).

| | **Displacement component** | **Proposed meshless method vs ABAQUS (C3D10)** |
|---|---|---|
| $L_\infty$ (in $m$) | x | $1.28 \times 10^{-3}$ |
| | y | $7.16 \times 10^{-4}$ |
| | z | $2.22 \times 10^{-3}$ |
| $L_{NRMSE}$ | x | $1.32 \times 10^{-2}$ |
| | y | $1.86 \times 10^{-2}$ |
| | z | $2.02 \times 10^{-2}$ |

Error norms reported in Table 15 indicate that the proposed meshless method gives, for practical purposes, equivalent results to the finite element method. For this example, it is difficult to judge whether meshless or finite element approach is more accurate.

Fig. 10, shows the deformation of the cylinder under indentation of 4 mm.

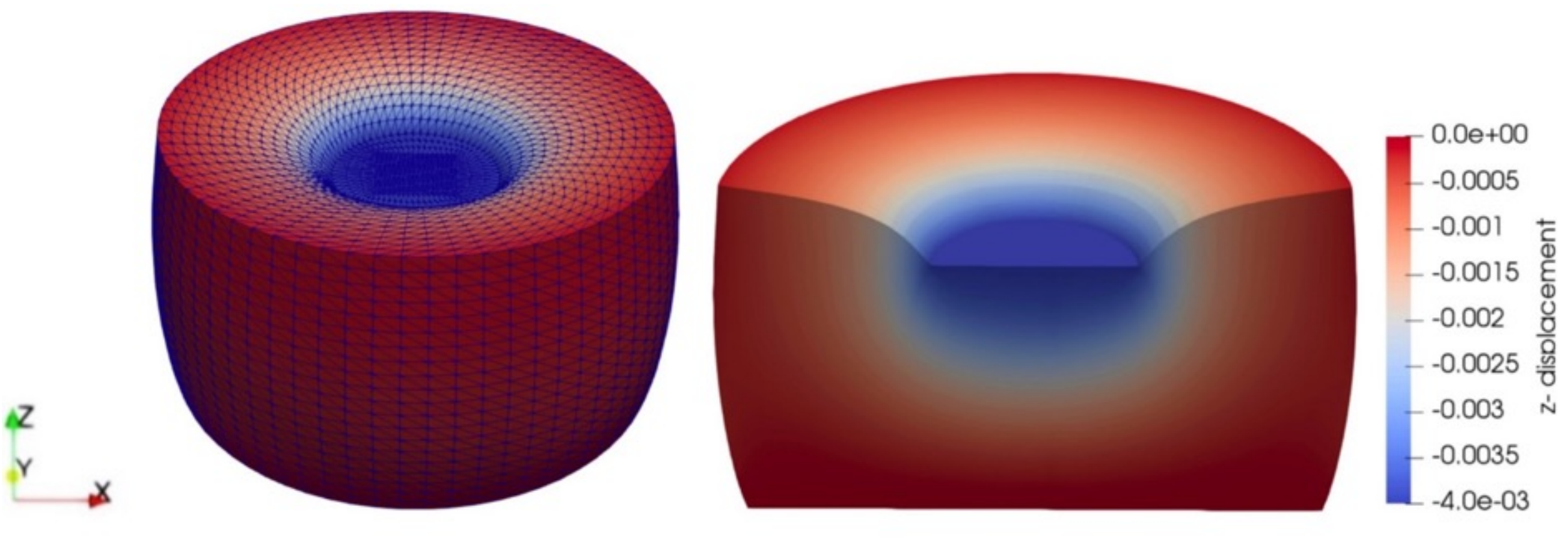

**(a)** **(b)**

**Fig. 10. (a)** Deformation field for the 4 mm (24% of the initial height) cylinder indentation and **(b)** cross section of the indented cylinder.

To demonstrate the accuracy of the essential boundary condition imposition, in Table 16 we list the $L_\infty^{(\boldsymbol{u},\boldsymbol{u}^h)}$ and the $L_2^{(\boldsymbol{u},\boldsymbol{u}^h)}$ error norms, computed at the rigidly constrained (green nodes in Fig. 9) and displaced nodes (red nodes in Fig. 9) of the indented cylinder.

**Table 16.** Maximum absolute difference $L_\infty^{(\boldsymbol{u},\boldsymbol{u}^h)}$ and root mean square error $L_2^{(\boldsymbol{u},\boldsymbol{u}^h)}$ between the fictitious ($\boldsymbol{u}$) and approximated ($\boldsymbol{u}^h$) computed at the rigidly constrained (green nodes in Fig. 9) and displaced nodes (red nodes in Fig. 9) of the indented cylinder.

| | **Displacement component** | **Rigidly constrained nodes** | **Displaced nodes** |
|---|---|---|---|
| $L_\infty^{(\boldsymbol{u},\boldsymbol{u}^h)}$ (in m) | $x$ | $8.12 \times 10^{-11}$ | $4.23 \times 10^{-11}$ |
| | $y$ | $3.54 \times 10^{-11}$ | $6.78 \times 10^{-11}$ |
| | $z$ | $4.26 \times 10^{-11}$ | $4.63 \times 10^{-11}$ |
| $L_2^{(\boldsymbol{u},\boldsymbol{u}^h)}$ (in m) | $x$ | $1.21 \times 10^{-11}$ | $8.22 \times 10^{-12}$ |
| | $y$ | $5.76 \times 10^{-12}$ | $6.68 \times 10^{-12}$ |
| | $z$ | $7.86 \times 10^{-12}$ | $9.46 \times 10^{-12}$ |

Now we demonstrate the applicability of the proposed method for very large indentation depths. Using our meshless algorithm, we were able to obtain robust results for the indentation depth of 12 mm corresponding to 70% of the initial height of the sample (Fig 11). ABAQUS failed to obtain results for such indentation depths. FEM solution fails (i.e. stops to converge - we selected $10^{-8}$ iteration step as the limit) due to large element distortion already at 7.5 mm indentation depth. To the best of our knowledge, the results given in Fig. 11 would be difficult to replicate with any other numerical method (without costly remeshing).

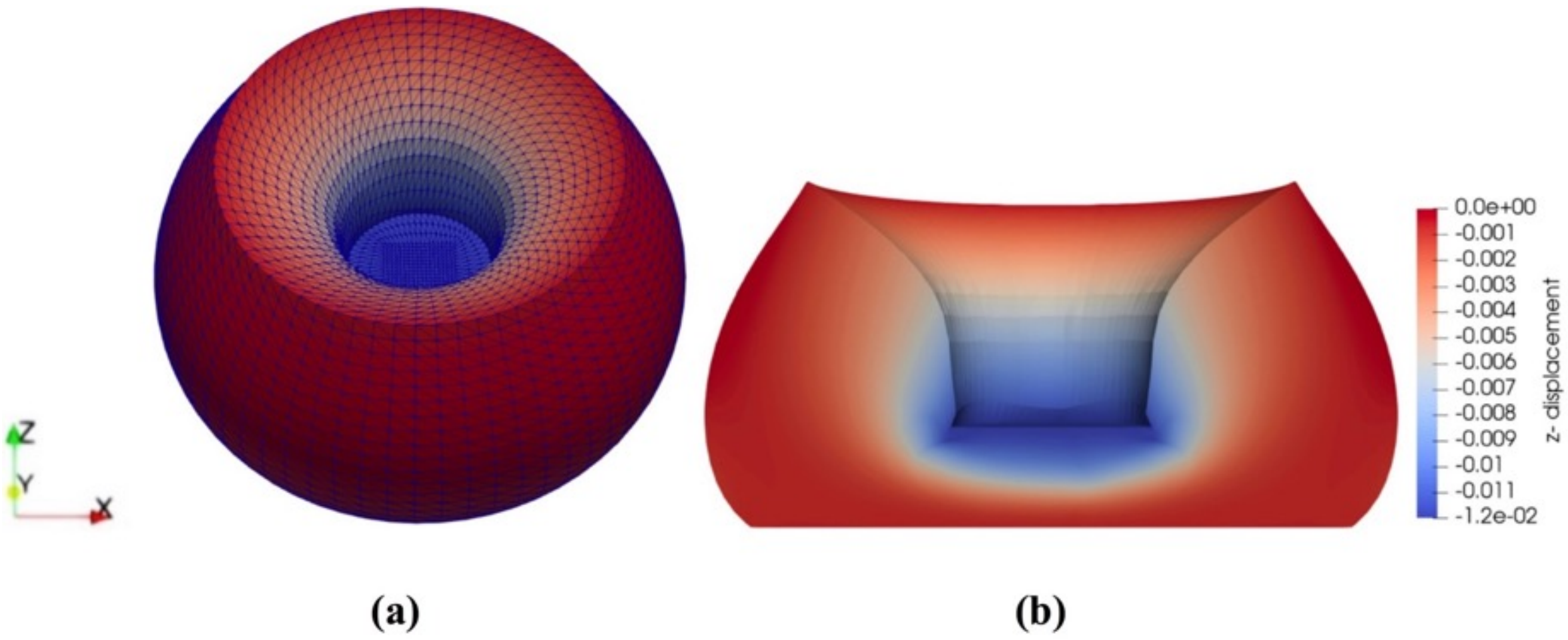

**Fig. 11. (a)** Deformation for the 12 mm cylinder indentation (~70% of the sample initial height) and **(b)** cross section of the indented cylinder.

*4.3 Brain indentation*

To demonstrate applicability of our methods to problems with complicated geometry we consider the patient-specific brain geometry, obtained from the magnetic resonance images (MRIs) acquired at Harvard Medical School and used previously in [7, 27]. The brain was discretized using 36,061 nodes and 198,061 integration cells with the key anatomical structures (the brain parenchyma, tumor and ventricles) included in the model (Fig. 12). It is worth noting that the generation of the meshless grid took seconds, while the construction of brain finite element hexahedral meshes takes many days of highly qualified analyst effort [28].

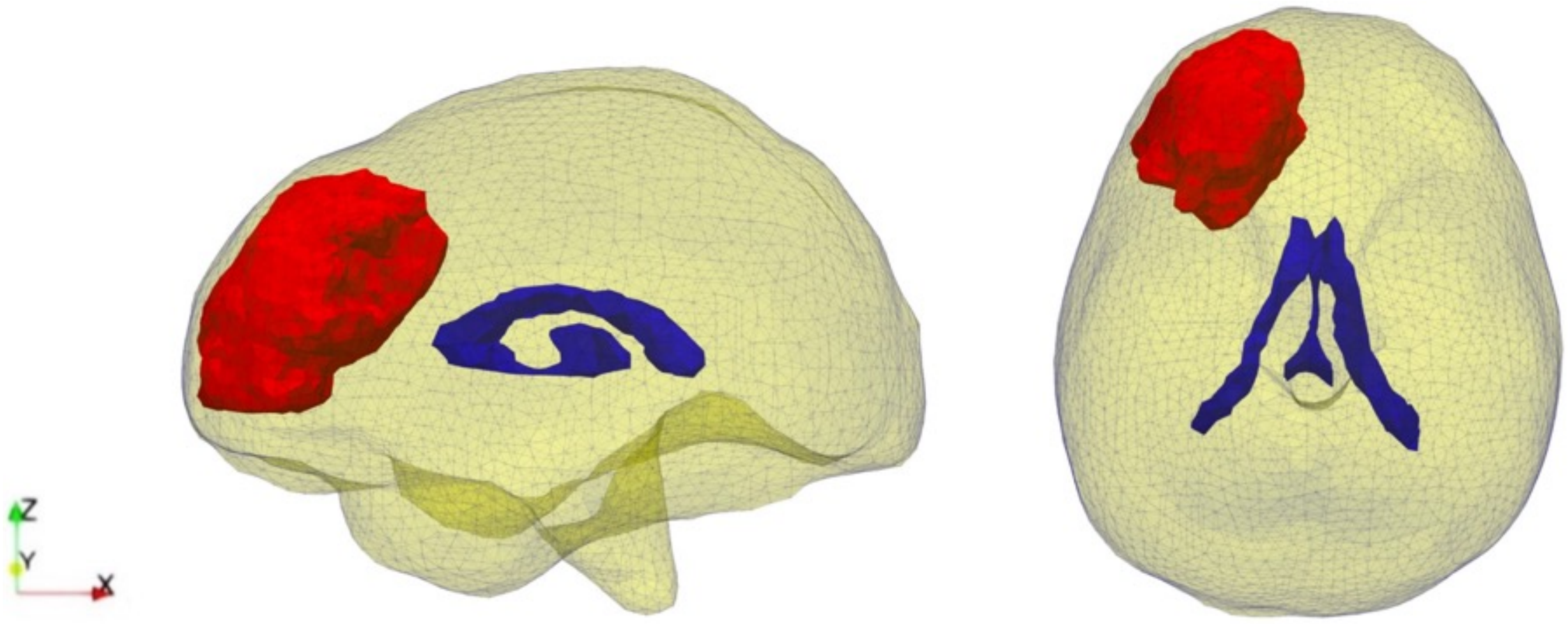

**Fig. 12. (a)** Side and **(b)** top view of surface visualizations of the analyzed continuum (human brain) geometry: parenchyma (yellow), tumor (red) and ventricles (blue).

To simulate brain indentation, the load was defined by prescribing displacements on the selected nodes of the external surface of the brain parenchyma (see Fig. 13). The displacement for each of these nodes was designated a value of 25 mm, and was smoothly imposed using a 3-4-5 polynomial, guaranteeing zero acceleration at the beginning and the end of load application. The nodes on the bottom (inferior) surface of the brain were rigidly constrained (Fig. 13). The material properties used for the different anatomical components of the brain were taken from [8] and are reported in Table 17.

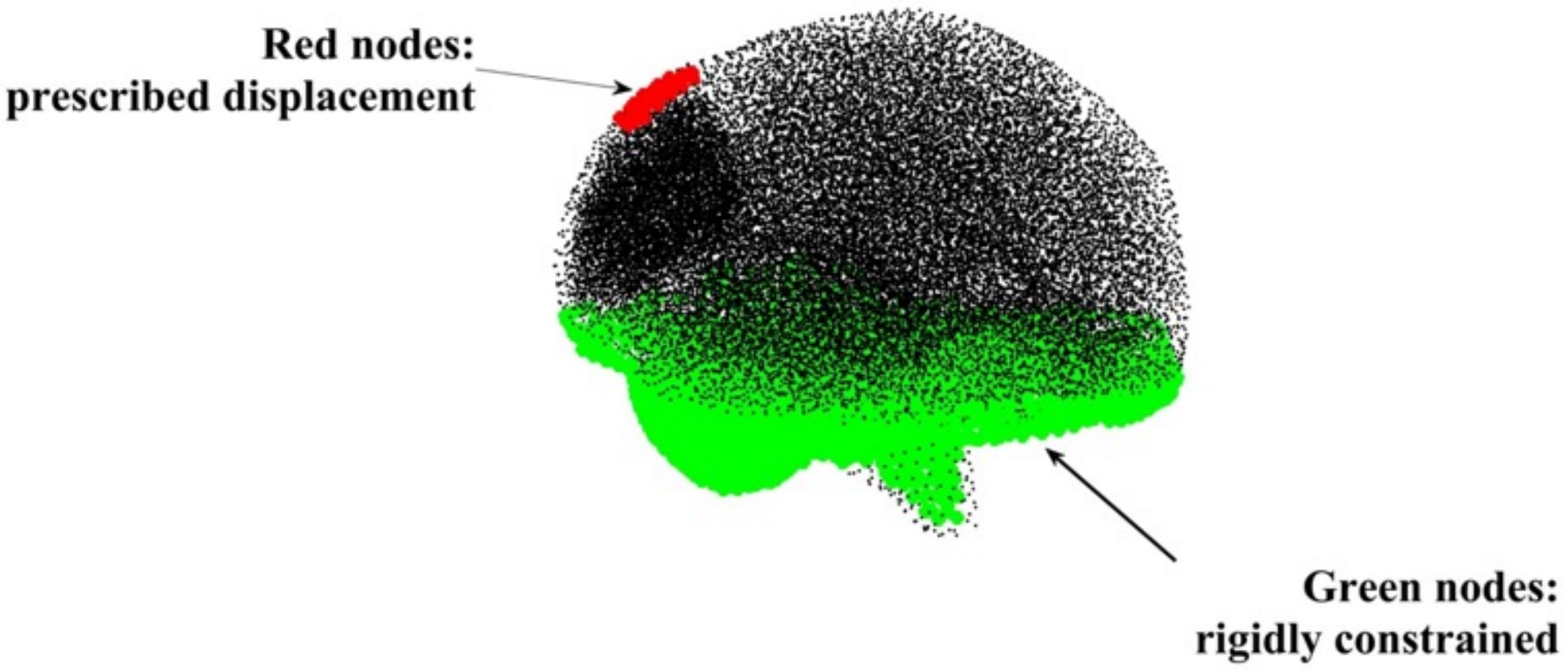

**Fig. 13.** Simulation of indentation of the brain. Meshless model consisting of 36,061 nodes and 198,061 integration cells in undeformed configuration. The prescribed displacement was applied at the nodes shown in red color. The nodes shown in green were rigidly constrained.

**Table 17.** Material properties for each component in the simulation of the brain indentation.

| Parameter | Parenchyma | Tumor | Ventricles |
|---|---|---|---|
| Density $\rho$ $(\mathrm{Kg/m^3})$ | 1,000 | 1,000 | 1,000 |
| Poisson ratio $v$ | 0.49 | 0.49 | 0.1 |
| Young modulus $E(Pa)$ | 3,000 | 6,000 | 10 |

We present results for indentation by 25 mm (Fig 14). They confirm the applicability of our method to domains with complex geometries. Table 18 lists the $L_\infty^{(\boldsymbol{u},\boldsymbol{u}^h)}$ and the $L_2^{(\boldsymbol{u},\boldsymbol{u}^h)}$ error norms, computed at the rigidly constrained (green nodes in Fig. 13) and displaced nodes (red nodes in Fig. 14).

**Table 18.** Maximum absolute difference $L_\infty^{(\boldsymbol{u},\boldsymbol{u}^h)}$ and root mean square error $L_2^{(\boldsymbol{u},\boldsymbol{u}^h)}$ between the fictitious $(\boldsymbol{u})$ and approximated $(\boldsymbol{u}^h)$ computed at the rigidly constrained (green nodes in Fig. 13) and displaced nodes (red nodes in Fig. 13) of the indented brain geometry.

| | Displacement component | Rigidly constrained nodes | Displaced nodes |
|---|---|---|---|
| $L_\infty^{(\boldsymbol{u},\boldsymbol{u}^h)}$ (in m) | $x$ | $5.99 \times 10^{-11}$ | $1.51 \times 10^{-11}$ |
| | $y$ | $3.90 \times 10^{-11}$ | $2.10 \times 10^{-11}$ |
| | $z$ | $1.93 \times 10^{-11}$ | $3.27 \times 10^{-11}$ |
| $L_2^{(\boldsymbol{u},\boldsymbol{u}^h)}$ (in m) | $x$ | $4.19 \times 10^{-12}$ | $4.15 \times 10^{-12}$ |
| | $y$ | $3.49 \times 10^{-12}$ | $5.20 \times 10^{-12}$ |
| | $z$ | $1.04 \times 10^{-12}$ | $9.64 \times 10^{-12}$ |

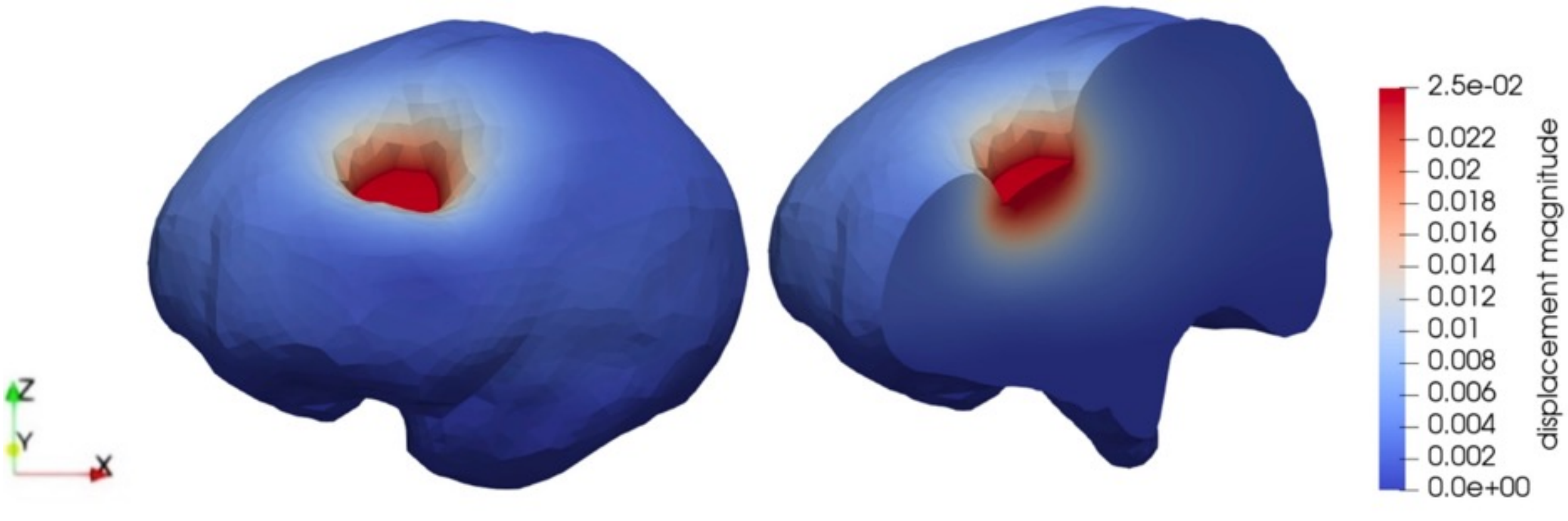

(a) (b)

**Fig. 14. (a)** Deformation for the 0.025 m brain geometry indentation and **(b)** cross section of the indented brain geometry.

Fig. 15 shows the undeformed and deformed geometry for the brain indentation example. We can observe the deformation of the tumor due to the indentation of the parenchyma.

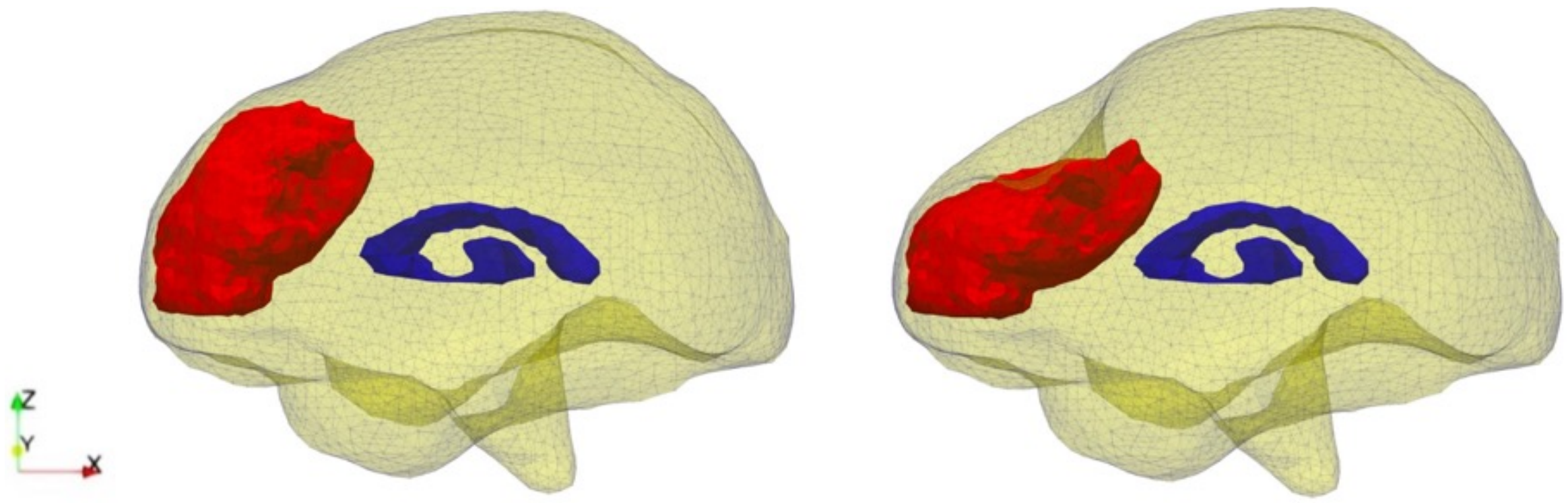

(a) (b)

**Fig. 15.** Side view of surface visualizations of the brain geometry; parenchyma (yellow), tumor (red) and ventricles (blue) are shown in the **(a)** undeformed and **(b)** deformed (due to indentation) state.

## 5. Conclusions

In this contribution, we introduced a simple, robust and accurate element-free Galerkin method. Its simplicity and robustness stems from the use of total Lagrangian formulation with explicit time integration, together with interpolating shape functions. Total Lagrangian formulation allows computation of shape functions and their derivatives only once, in the pre-processing stage. Explicit time integration circumvents problems of equation nonlinearities and entirely avoids any algebraic equation solving. Model stiffness and tangent matrices do not need to be assembled and entire solution progresses in a degree-of-freedom-by-degree-of-freedom manner. The (conditional) stability of the scheme is guaranteed by appropriate selection of a time step [29]. Accurate imposition of essential boundary conditions is often considered difficult using a meshless approach. Nevertheless, our approach allows direct imposition of essential boundary conditions because of the interpolating nature of our shape functions. This is a decisive advantage of the proposed approach over the previously published MTLED algorithm [7] that requires very sophisticated and difficult to implement algorithm for essential boundary condition imposition [30].

We demonstrated the convergence and accuracy of the proposed method on a variety of simple and complicated 3D examples. We believe that some of the presented numerical results would be very difficult to obtain with any other numerical method.

An obvious limitation of the proposed approach is the use of a simple Gauss scheme for numerical integration. Theoretically, this approach is not satisfactory because exact integration of rational functions cannot be achieved with this method. However, our results demonstrate that overall accuracy of our simulations is satisfactory. The alternative would be to use an adaptive integration scheme with guaranteed accuracy, such as e.g. in [31]. However, using such a complicated approach, while theoretically preferable, would undermine the simplicity and robustness of the method and especially of the code.

**Acknowledgements**

This research was supported in part by the Australian Government through the Australian Research Council's Discovery Projects funding scheme (project DP160100714). The views expressed herein are those of the authors and are not necessarily those of the Australian Research Council.

## References

1. Atluri, S.N., *The Meshless Method*. 2002: Tech. Sci. Press, Forsyth.
2. Atluri, S.N. and T. Zhu, *A new Meshless Local Petrov-Galerkin (MLPG) approach in computational mechanics.* Computational Mechanics, 1998. **22**(2): p. 117-127.
3. Monaghan, J.J., *Smoothed Particle Hydrodynamics.* Annual Review of Astronomy and Astrophysics, 1992. **30**(1): p. 543-574.
4. Belytschko, T., Y.Y. Lu, and L. Gu, *Element-free Galerkin methods.* International Journal for Numerical Methods in Engineering, 1994. **37**(2): p. 229-256.
5. Almasi, A., Beel, A., Kim, T.-Y., Michopoulos, J. G., *Strong-Form Collocation Method for Solidification and Mechanical Analysis of Polycrystalline Materials.* Journal of Engineering Mechanics, 2019. **145**(10): p. 04019082.
6. Bourantas, G.C., Mountris, K.A., Loukopoulos, V.C., Lavier, L., Joldes, G.R., Wittek, A., Miller, K., *Strong-form approach to elasticity: Hybrid finite difference-meshless collocation method (FDMCM).* Applied Mathematical Modelling, 2018. **57**: p. 316-338.
7. Joldes, G.R., Bourantas, G.C., Zwick, B.F., Chowdhury, H., Wittek, A., Agrawal, S. Mountris, K.A., Hyde, D., Warfield, S.K., Miller, K., *Suite of meshless algorithms for accurate computation of soft tissue deformation for surgical simulation.* Medical Image Analysis, 2019. **56**: p. 152-171.
8. Miller, K., Horton, A., Joldes, G.R., Wittek, A., *Beyond finite elements: A comprehensive, patient-specific neurosurgical simulation utilizing a meshless method.* Journal of Biomechanics, 2012. **45**(15): p. 2698-2701.
9. Wittek, A., Bourantas, G. C., Zwick, B.F., Joldes, G.R., Estebban, L., Miller, K., *Mathematical Modeling and Computer Simulation of Needle Insertion into Soft Tissue*. 2020.
10. Joldes, G.R., Chowdhury, H.A., Wittek, A., Doyle, B., Miller, K., *Modified moving least squares with polynomial bases for scattered data approximation.* Applied Mathematics and Computation, 2015. **266**: p. 893-902.
11. Joldes G.R., Teakle, P., Wittek, A., Miller, K., *Computation of accurate solutions when using element-free Galerkin methods for solving structural problems.* Engineering Computations, 2017. **34**(3): p. 902-920.
12. Bathe, K.-J., *Finite Element Procedures*. 1996, New Jersey: Prentice-Hall.
13. Miller, K., Joldes, G.R., Lance, D., Wittek, A., *Total Lagrangian Explicit Dynamics Finite Element Algorithm for Computing Soft Tissue Deformation.* Communications in Numerical Methods in Engineering, 2007. **23**(2): p. 121-134.
14. Horton, A., Wittek, A., Joldes, G.R., Miller, K., *A meshless Total Lagrangian explicit dynamics algorithm for surgical simulation.* International Journal for Numerical Methods in Biomedical Engineering, 2010. **26**(8): p. 977-998.
15. Joldes, G.R., A. Wittek, and K. Miller, *Computation of intra-operative brain shift using dynamic relaxation.* Computer Methods in Applied Mechanics and Engineering, 2009. **198**(41): p. 3313-3320.
16. Joldes, G.R., A. Wittek, and K. Miller, *An adaptive dynamic relaxation method for solving nonlinear finite element problems. Application to brain shift estimation.* International Journal for Numerical Methods in Biomedical Engineering, 2011. **27**(2): p. 173-185.
17. Liu, G.R., *Mesh Free Methods: Moving Beyond the Finite Element Method.* 2003, Boca Raton: CRC Press.
18. Lancaster, P. and K. Salkauskas, *Surfaces generated by moving least-squares methods.* Mathematics of Computation, 1981. **37**(155): p. 141-158.

19. Most, T. and C. Bucher, A moving least squares weighting function for the element-free Galerkin method which almost fulfills essential boundary conditions. 2005. Structural Engineering and Mechanics **21(3)** 315-332.
20. *ABAQUS, I., 2014. ABAQUS Online Documentation: Version 6.14.*
21. Holzapfel, G.A., *Nonlinear solid mechanics: a continuum approach for engineering*. 2000, New York: Wiley.
22. *Ogden, R. W., Non-Linear Elastic Deformations. Ellis Horwood Ltd. 1984. Chichester, Distributors: John Wiley & Sons Ltd.*
23. Miller, K., *Method of testing very soft biological tissues in compression.* Journal of Biomechanics, 2005. **38**(1): p. 153-158.
24. Morriss, L., A. Wittek, and K. Miller, *Compression testing of very soft biological tissues using semi-confined configuration—A word of caution.* Journal of Biomechanics, 2008. **41**(1): p. 235-238.
25. *FEniCS Project* https://fenicsproject.org/. 2020.
26. *Hyperelasticity* https://fenicsproject.org/docs/dolfin/latest/python/demos/hyperelasticity/demo_hyperelasticity.py.html. 2020.
27. Wittek, A., Joldes, G.R., Couton, M., Warfield, S.K., Miller, K., *Patient-specific non-linear finite element modelling for predicting soft organ deformation in real-time; Application to non-rigid neuroimage registration.* Progress in Biophysics and Molecular Biology, 2010. **103**: p. 292-303.
28. Wittek, A., Grosland, N.M., Joldes, G.R., Magnotta, V., Miller, K., *From finite element meshes to clouds of points: A review of methods for generation of computational biomechanics models for patient-specific applications.* Annals of Biomedical Engineering, 2016. **44**(1): p. 3-15.
29. Joldes, G.R., A. Wittek, and K. Miller, *Stable time step estimates for mesh-free particle methods.* International Journal for Numerical Methods in Engineering, 2012. **91**(4): p. 450-456.
30. J Joldes, G.R., Chowdhury, H., Wittek, A., Miller, K., *A new method for essential boundary conditions imposition in explicit meshless methods.* Engineering Analysis with Boundary Elements, 2017. **80**: p. 94-104.
31. Joldes, G.R., A. Wittek, and K. Miller, *Adaptive numerical integration in Element-Free Galerkin methods for elliptic boundary value problems.* Engineering Analysis with Boundary Elements, 2015. **51**: p. 52-63.